\documentclass[11pt]{article}

\newcommand{\be}{\begin{equation}}
\newcommand{\ee}{\end{equation}}
\usepackage{amsmath}
\usepackage{amssymb}
\usepackage{latexsym}
\usepackage{epsfig}

\newcommand{\Z}{\mathbb{Z}}
\newcommand{\R}{\mathbb{R}}

\newcommand{\e}{{\check e}}

\newcommand{\bea}{\begin{eqnarray}}
\newcommand{\eea}{\end{eqnarray}}

\newcommand{\RRR}{{\hbox{\rm R\kern-2.35mm R}}}

\def\ZZZ{{\hbox{ Z\kern-1.6mm Z}}}

\begin{document}

\begin{titlepage}
\rightline{March 24, 2010}
%\rightline{\tt arXiv:xxxx.xxxx}
\rightline{\tt  Imperial-TP-2010-CH-01}
\rightline{\tt MIT-CTP-4128}
\begin{center}
\vskip 2.5cm
{\Large \bf {
Background independent action for double field theory
}}\\
\vskip 1.0cm
{\large {Olaf Hohm${}^1\hskip-4pt ,$ Chris Hull${}^2\hskip-4pt ,$ and Barton Zwiebach${}^1$}}
\vskip 0.5cm
{\it {${}^1$Center for Theoretical Physics}}\\
{\it {Massachusetts Institute of Technology}}\\
{\it {Cambridge, MA 02139, USA}}\\
ohohm@mit.edu, zwiebach@mit.edu
\vskip 0.7cm
{\it {${}^2$The Blackett Laboratory}}\\
{\it {Imperial College London}}\\
{\it {Prince Consort Road}}\\
{\it { London SW7 @AZ, U.K.}}\\
c.hull@imperial.ac.uk

\vskip 1.5cm
{\bf Abstract}
\end{center}

\vskip 0.5cm

\noindent
\begin{narrower}
Double field theory describes a massless subsector of closed string theory with both momentum and winding excitations.  
The gauge algebra is governed by the Courant bracket in 
certain subsectors of this double field theory. 
We construct the associated nonlinear background-independent 
action that is T-duality invariant  and realizes the Courant gauge algebra. 
 The action is the sum of  a standard action for gravity, antisymmetric tensor, and dilaton fields written with ordinary derivatives,
 a similar action for dual fields with dual derivatives, 
 and a mixed term that is needed for gauge invariance.

\end{narrower}

\end{titlepage}

\newpage

\tableofcontents
\baselineskip=16pt

\section{Introduction}

Double field theory is a field theory on the doubled torus that arises when the
usual  toroidal coordinates $x^i$
are supplemented with coordinates $\tilde x_i$ associated
with winding excitations.  Closed string field theory in toroidal backgrounds  is, by construction, a double field theory~\cite{Kugo:1992md}.  In recent papers we have
started the construction of a double field theory based on the massless fields of closed string theory~\cite{Hull:2009mi,Hull:2009zb}.  In this theory the gravity field $h_{ij}$, the
antisymmetric tensor $b_{ij}$, and the dilaton $d$ all depend on the coordinates $x^i$ and $\tilde x_i$. The construction is novel and requires a
constraint that arises from the $L_0-\bar L_0=0$ constraint of
closed string theory:
all fields and gauge parameters must be annihilated by the
differential operator $\partial_i \tilde \partial^i$, where a sum over $i$ is
understood.  This `massless' double field theory was constructed to cubic order in the
fields and to this order it is gauge invariant and
has a remarkable  T-duality symmetry.
It is an open question
whether a complete nonlinear gauge-invariant extension of this massless theory exists.
It would not be a
conventional low-energy limit of closed string theory,
and it is not yet clear whether or not this could be a consistent 
truncation of string theory.
It would be
 simpler than closed string theory, however,  
and
the ideas of doubled geometry   as well as 
  some of the essential features of closed string theory
should   be more accessible there.  Earlier
work in double field theory includes that of
Tseytlin~\cite{Tseytlin:1990nb} and   Siegel~\cite{Siegel:1993th}.

Rather than proceed with the
full higher order construction, we
will here focus on a subsector of the full double field theory satisfying a
constraint that is stronger than the one discussed above and complete the construction of this subsector.
The constraint from string field theory requires that all fields and gauge parameters
are annihilated by
$\partial_i \tilde \partial^i$, but the strong constraint requires that in addition
 all {\it  products} of  fields and gauge parameters are also annihilated by $\partial_i \tilde \partial^i$.
With this strong  constraint we were
able~\cite{Hull:2009zb} to find a consistent set
of gauge transformations to all orders and
 show  that their gauge
algebra is governed by the Courant bracket of generalised
geometry~\cite{Tcourant,Hitchin,Gualtieri,LWX,Gualtieri2}.  In this
paper we continue the analysis and construct the action to all orders in the fluctuating fields. We do so  by constructing a background independent
action.  The fluctuations and the background fields of the cubic action
give enough clues for the definition of full fields in terms of which the
action can be neatly constructed.  We use the field ${\cal E}_{ij} = g_{ij}+ b_{ij}$, 
comprising of the metric $g_{ij}$ and the antisymmetric
tensor $b_{ij}$, as well as a dilaton field $d$.

A striking feature of our theory is its duality symmetry.
If the spacetime is the product of $n$-dimensional Minkowski space
$\mathbb{R}^{n-1,1}$ and a torus $T^d$, string theory has  $O(d,d;\Z)$ 
T-duality symmetry %oh 
\cite{Giveon:1988tt,Shapere:1988zv} (see also \cite{Giveon:1994fu}
and references therein).
In our formulation,  the torus coordinates are doubled, and the double field theory has an $O(d,d;\Z)$ symmetry that acts naturally on the $2d$ coordinates $(x, \tilde x)$  of the doubled torus.
It is formally useful 
to double the non-compact coordinates also, and our double field theory
in that case has a continuous $O(n,n)$ symmetry.
In particular, when formulated in non-compact $\mathbb{R}^{2D}$, 
the double field theory has a continuous $O(D,D)$ symmetry.
Compactifying $2d$ of these dimensions on a double torus breaks this to
$O(n,n)\times O(d,d;\Z)$.
Restricting  the fields to be independent of the extra $n$ non-compact coordinates breaks $O(n,n)$ to the Lorentz group $O(n-1,1)$.
Then this $O(D,D)$ symmetry
ensures the Lorentz and T-duality symmetries of the compactified cases relevant to string theory, and it will often
be convenient to simply refer to the $O(D,D)$ symmetry in what follows.
The constraint $\partial_i \tilde \partial^i=0$ is $O(D,D)$ invariant
and can be written as
$\partial^M \partial_M=0$
upon collecting the coordinates $x^i$ and
dual coordinates
$\tilde x_i$ into an $O(D,D)$ vector $X^M$
and using the constant $O(D,D)$ invariant metric $\eta_{MN}$ to raise and lower indices, so that
$\partial^M \partial_M=\eta ^{MN}\partial_M \partial_N$.

 The strong constraint  introduced above 
  is in fact so strong that it means that
 there is a choice of coordinates $(x', \tilde x')$, related to the original coordinates
 by  $O(D,D)$, in which
the doubled
 fields only depend on half of the coordinates:  they depend on the $x'$ and
 are independent of the $\tilde x'$. We refer to such fields as restricted to the null subspace with coordinates $x'$~\cite{Hull:2009zb}.
We shall 
show that all solutions of the strong constraint are in fact related to a conventional field theory by
 $O(D,D)$ in this way.

It is interesting to compare with the work of Siegel~\cite{Siegel:1993th}, who performed
a direct construction of the strongly constrained theory discussed
above.  The gauge algebra, which coincides with ours, was his
starting point.  Siegel also constructed an action, but his formalism
is sufficiently different from ours  that direct comparison is not
straightforward. He uses vielbeins and an enlarged tangent space symmetry, while we simply
use the familiar field  ${\cal E}_{ij}$ and the dilaton.
We hope that our identification  
and discussion of
the role of the Courant bracket as well as the explicit and concrete
expressions for the action given in this paper will stimulate further
analysis and a detailed comparison.

The main result in this paper is the action, which takes the form
\bea
\label{THEActionINTRO}
 \begin{split}\hskip-10pt
  S \ = \ \int \,dx d\tilde{x}~
  e^{-2d}\Big[&
  -\frac{1}{4} \,g^{ik}g^{jl}   \,   {\cal D}^{p}{\cal E}_{kl}\,
  {\cal D}_{p}{\cal E}_{ij}
  +\frac{1}{4}g^{kl} \bigl( {\cal D}^{j}{\cal E}_{ik}
  {\cal D}^{i}{\cal E}_{jl}  + \bar{\cal D}^{j}{\cal E}_{ki}\,
  \bar{\cal D}^{i}{\cal E}_{lj} \bigr)~
\\ &    + \bigl( {\cal D}^{i}\hskip-1.5pt d~\bar{\cal D}^{j}{\cal E}_{ij}
 +\bar{{\cal D}}^{i}\hskip-1.5pt d~{\cal D}^{j}{\cal E}_{ji}\bigr)
 +4{\cal D}^{i}\hskip-1.5pt d \,{\cal D}_{i}d ~\Big]\;.
 \end{split}
 \eea
In this action, the calligraphic derivatives are defined by
\be
\label{groihffkdf}
{\cal D}_i \ \equiv \ {\partial\over \partial x^i} - {\cal E}_{ik} \,{\partial \over \partial\tilde x_k}\,,
~~~~\bar {\cal D}_i \ \equiv \ {\partial\over \partial x^i} + {\cal E}_{ki}\, {\partial \over \partial\tilde x_k}\,,
\ee
and all indices are raised with  $g^{ij}$, which is the inverse
of the metric $g_{ij} = {1\over 2} ({\cal E}_{ij} + {\cal E}_{ji})$.
The gauge transformations that leave the action invariant take the form
\begin{equation}
 \label{finalgtINTRO}
 \begin{split}
  \delta {\cal E}_{ij} \ &= \ {\cal D}_i\tilde{\xi}_{j}-\bar{{\cal D}}_{j}\tilde{\xi}_{i}
  +\xi^{M}\partial_{M}{\cal E}_{ij}
  +{\cal D}_{i}\xi^{k}{\cal E}_{kj}+\bar{\cal D}_{j}\xi^{k}{\cal E}_{ik}\;,\\[0.5ex]
 \delta d~ \ &= - {1\over 2}  \partial_M \xi^M + \xi^M \partial_M \,d\,.
 \end{split}
 \end{equation}
Here $\xi^M\partial_M = \xi^i \partial_i + \tilde \xi_i \tilde \partial^i$ and
$\partial_M\xi^M = \partial_i\xi^i  +\tilde \partial^i \tilde \xi_i $.
All fields, gauge parameters, and all possible products of them are
assumed to be annihilated by $\partial^M \partial_M$.  
The algebra of gauge transformations is governed by an $O(D,D)$
covariant  ``C-bracket''~\cite{Siegel:1993th,Hull:2009zb} which reduces 
to the 
Courant bracket when fields and gauge parameters  
are T-dual to ones that are  independent of winding coordinates.
This action is also invariant under the non-linearly realized $O(D,D)$ duality transformations that transform ${\cal E}$ projectively and leave
the dilaton invariant.  Using matrix notation,
 \begin{equation}
  {\cal E}^{\prime}(X^{\prime}) \ = \ (a{\cal E}(X)+b)(c{\cal E}(X)+d)^{-1}\;,
  \quad
  d^{\prime}(X^{\prime}) \ = \ d(X)\;, \quad X' = h X\,,
 \end{equation}
 where
  \begin{equation}
 h= \begin{pmatrix} a &   b \\ c & d \end{pmatrix} \ \in \ O(D,D)\;.
 \end{equation}

Finally,  when fields are assumed to be independent
of the $\tilde x$ coordinates, the action~(\ref{THEActionINTRO}) reduces to a form that is field-redefinition
equivalent to the familiar 
action
 \bea\label{original}
  S_* \ = \ \int dx \sqrt{-g}e^{-2\phi}\left[R+4(\partial\phi)^2-\frac{1}{12}H^2\right]\,.
 \eea
 In particular, the usual scalar dilaton  $\phi$ is related to the field $d$ used here by
 $\sqrt{-g} e^{-2\phi} = e^{-2d}$, so that $ e^{-2d}$ is a scalar density.
A by-product of our analysis is
that we find the explicit  form of the field
redefinitions that relate
the metric and $b$-field variables used in string theory to those used
in the conventional Einstein plus $b$-field theory (see \S\ref{BindGaug}), completing the work of Refs.~\cite{Michishita:2006dr,Hull:2009mi}.

There has been some interest in
gravity theories that replace the metric with a non-symmetric tensor field; see e.g.~\cite{Damour:1992bt}. The field ${\cal E}_{ij}$ is non-symmetric, and we believe (\ref{THEActionINTRO}) is a natural
action written in terms of this variable.
After rotating to a T-duality frame in which all tilde-derivatives
are zero, the
action (given in (\ref{THEActionvm1})) is a rather simple rewriting of the conventional action $S_*$ using the 
${\cal E}$ field  
variable. 
The theory defined by (\ref{THEActionINTRO}) provides therefore a
reformulation of $S_*$ in which $O(D,D)$ is a symmetry. Moreover, in contrast to theories
of non-symmetric gravity of the type discussed in~\cite{Damour:1992bt}, here the symmetric
and antisymmetric parts of ${\cal E}_{ij}$ do not provide irreducible representations
of the gauge and duality symmetries
but rather transform into each other.

All terms in the action~(\ref{THEActionINTRO}) contain two derivatives.
It is
natural to examine a  tilde-derivative ($\tilde \partial$) expansion
$S = S^{(0)} + S^{(1)} + S^{(2)}$ where $S^{(k)}$ contains $k$ tilde derivatives.   The action $S^{(0)} ({\cal E}, d, \partial)$ is obtained
by setting all terms with
 tilde-derivatives to zero and thus, as mentioned above, is
equivalent to $S_*$.   It is interesting that $S^{(2)}$ is in fact equal to
$S^{(0)} ({\cal E}^{-1}, d, \tilde \partial)$, the conventional action
for gravity, $b$-field, and dilaton but with ${\cal E}\to {\cal E}^{-1}$ and
$\partial \to \tilde \partial$, the changes associated with inversion
duality.  The action $S^{(1)}$ is a curious mix, with terms that involve
one derivative of each type.   Gauge invariance of the action
will  be demonstrated
 explicitly
 using a similar tilde-derivative expansion of the
gauge transformations.
Specifically, the gauge transformations (\ref{finalgtINTRO}) can be rewritten
as
  \bea\label{gaugeexpand}
   \delta_{\xi}{\cal E}_{ij} &=& {\cal L}_{\xi}{\cal E}_{ij}
   +\partial_{i}\tilde{\xi}_{j}-\partial_{j}\tilde{\xi}_{i} \\[0.5ex] \nonumber
   &&\hskip-10pt+\,{\cal L}_{\tilde{\xi}}{\cal E}_{ij}-{\cal E}_{ik}
   \left(\tilde{\partial}^{k}\xi^{l}-\tilde{\partial}^{l}\xi^{k}\right){\cal E}_{lj}\;.
  \eea
Here, ${\cal L}_{\xi}$ and ${\cal L}_{\tilde{\xi}}$ are the Lie derivatives with respect to 
the parameters $\xi^i$ and $\tilde\xi_i$ respectively.
We note that, despite
appearance in (\ref{finalgtINTRO}), these two gauge parameters enter completely democratically
as they combine naturally into the $O(D,D)$ vector $\xi^{M}=(\tilde{\xi}_{i},\xi^{i})$.
In fact, we will show that under the inversion duality ${\cal E}\to {\cal E}^{-1}$,
$\partial \to \tilde \partial$ the role of $\xi^{i}$ and $\tilde{\xi}_{i}$ in
(\ref{gaugeexpand}) gets precisely interchanged.

We give some preliminary discussion  
of an  attempt to formulate  an
`$O(D,D)$ geometry'.  The analysis of $O(D,D)$ invariance of the
action and covariance of the gauge transformations is facilitated
by the introduction of $O(D,D)$ `covariant' derivatives that map
$O(D,D)$ tensors to $O(D,D)$ tensors.  We also construct
an $O(D,D)$-invariant scalar curvature ${\cal R}({\cal E}, d)$,
all of whose terms contain two derivatives (see 
eqn.~(\ref{generalscalar})).
Being a scalar means that it transforms under gauge transformations
as $\delta_{\xi}{\cal R}=\xi^{M}\partial_{M}{\cal R}$.  Since $e^{-2d}$ is a density
we obtain an $O(D,D)$ invariant and gauge invariant action by
writing
 \bea\label{masteractionINTRO}
  S' \ = \ \int dxd\tilde{x}\,e^{-2d}\,{\cal R}({\cal E},d)\;.
 \eea
The two actions $S'$ and $S$ in
(\ref{THEActionINTRO}) differ by
integrals of total derivatives and so are equivalent.
It is a nontrivial fact that the dilaton equation
of motion following from (\ref{masteractionINTRO}) is simply
${\cal R} =0$.  The variations of $d$ within ${\cal R}$ combine to a
total derivative and do not contribute to the field equation.  When we
set tilde-derivatives to zero we find that,
after the field redefinition
that trades $d$ for $\phi$, the scalar ${\cal R}$ becomes
 \bea\label{EinsteinscalarINTRO}
  {\cal R}({\cal E},d)\Big|_{\tilde{\partial}=0} \ = \
  R+4\square\phi-4(\partial\phi)^2-\frac{1}{12}H^2\;.
 \eea
Thus assuming that all fields are independent of the $\tilde x$
coordinates the action $S'$ (without discarding total
derivatives) becomes:
 \bea\label{original11}
  S'_*
   \ = \ \int dx \sqrt{-g}e^{-2\phi}\left[R+4\square\phi-4(\partial\phi)^2-\frac{1}{12}H^2\right]\,.
 \eea
This action inherits the curious property of $S'$:
the equation
of motion for the dilaton is precisely the vanishing of the terms
in the square
brackets. Note that this is not true for the action $S_*$ in (\ref{original}).
The curvature ${\cal R}$ is precisely the beta function $\beta^d$ for the 
$O(D,D)$  scalar
dilaton $d$. 
The spacetime action
in~\cite{Tseytlin:1988rr} used a Lagrangian  proportional to $\beta^d$. 
The gauge scalar ${\cal R}({\cal E}, d)$ is
also an $O(D,D)$ scalar and can be viewed as a
generalization of the scalar
curvature of Einstein's theory.  It is interesting that the object that
admits a generalization is the precise combination of terms indicated
in (\ref{EinsteinscalarINTRO}). 

We also investigate the relation between the $O(D,D)$ duality symmetry and the gauge symmetry.
We show that $GL(D)$ transformations and constant shifts of the $b$-field are special gauge symmetries, but in general the remaining $O(D,D)$ transformations do not arise from gauge symmetries. However, we find that in the special case in which we truncate the theory to one with fields that are independent of $d$ coordinates 
and their  $d$ duals
there is an $O(d,d)$ symmetry that arises from gauge symmetries.
This gives a geometric insight into the $O(d,d)$ symmetry \cite{Maharana:1992my} that emerges from  Kaluza-Klein reduction on a $d$-torus.

\section{Action and gauge transformations}\setcounter{equation}{0}

In this section we start by proving background independence of the cubic action given in
\cite{Hull:2009mi}.  Then we determine the manifestly background independent form both for the action and the 
gauge transformations 
 to all orders in the fluctuations. 
In addition, 
we prove the $O(D,D)$ invariance of the action and
the $O(D,D)$ covariance of the gauge transformations. Finally, 
we show that  the strong form of the constraint 
$\partial^{M}\partial_{M}=0$ guarantees that
one can always find an $O(D,D)$ transformation that rotates 
to a duality frame where all fields are independent of the winding coordinates.

\subsection{Background independence}

In~\cite{Hull:2009mi} we constructed the double field theory action
for the dilaton $d$ and the field $e_{ij}$
giving the fluctuation in the metric and anti-symmetric tensor gauge field
around the constant
background $E_{ij} = G_{ij} + B_{ij}$.  The action was obtained to cubic order in the fields $e_{ij}$ and $d$ and takes the form (with
$2\kappa^2=1$):
\be
\label{redef-action}
\begin{split}
S &=    \hskip-2pt\int [d x d\tilde x] \,
\Bigl[\, \, {1\over 4} e_{ij} \square e^{ij}  + {1\over 4} (\bar D^j e_{ij})^2
+ {1\over 4} ( D^i e_{ij})^2  - 2 \, d \, D^i \bar D^j e_{ij}
\,-\, 4 \,d \, \square \, d~~~
\\[0.6ex]
&\hskip15pt + {1\over 4} \,e_{ij}\Bigl( \,(D^i e_{kl})  ( \bar D^j e^{kl} )
 -(D^ie_{kl} )\,( \bar D^l e^{kj})
 - (D^k e^{il} )( \bar D^j e_{kl} )  \Bigr)~
\\[1.0ex]
&\hskip15pt +
{1\over 2} d\,
\Bigl(  (D^ie_{ij})^2  + (\bar D^j e_{ij})^2
+{1\over 2}(D_k e_{ij})^2
 +{1\over 2}( \bar D_k e_{ij} )^2
 + 2e^{ij} (D_i   D^k e_{kj}  + \bar D_j  \bar D^k e_{ik} )
   \Bigr)~  \\[0.9ex]
&\hskip15pt +
4\,e_{ij} d\, D^i \bar D^jd
 + 4 \,d^2 \,\square\, d~\Bigr]\,.
 \end{split}
\ee
In~\cite{Hull:2009mi} the corresponding gauge transformations were found to linear order in the fields and
this action was shown to be invariant under these   gauge transformations up to terms quadratic in the fields, provided
 the fields and gauge parameters are 
 annihilated by $\partial_{i}\tilde{\partial}^{i}$. 
The background dependence in the action
enters through the derivatives
\be
\label{groihfgruu8774}
D_i \ = \ {\partial\over \partial x^i} - E_{ik} \,{\partial \over \partial\tilde x_k}\,,
~~~~\bar D_i \ = \ {\partial\over \partial x^i} + {E}^t_{ik}\, {\partial \over \partial\tilde x_k}\,,
\ee
and the inverse metric $G^{ij}$, which is used to raise indices in this subsection.
Note that the field $e_{ij}$ and the derivatives $D_i$ and $\bar D_i$ are defined with indices down.

Background independence
 is a property that
the above action should inherit from the full string theory.   This means that one can absorb a
constant part of the fluctuation field $e_{ij}$ into
a change of the background field $E_{ij}$.
The dilaton plays no role in the background dependence; we
will display only the $E$ and $e$ dependence in the action
and write
$S[E_{ij}, e_{ij}]$. Let $\chi_{ij}$ be an infinitesimal, constant part
of the field $e_{ij}$. The statement
of background independence is that
\be
 S\big[E_{ij}\, ,\;e_{ij}+\chi_{ij} \big] \ = \ S\big[E_{ij}+\chi_{ij}\, ,\;e'_{ij}
 = e_{ij} + f_{ij}(\chi, e) \big]\;,
\ee
where $f(\chi,e)$ is a function that is linear in $\chi$ and to leading order
linear in $e_{ij}$.  This function is needed in general; it means that the new fluctuation field $e'_{ij}$ is a field redefined version of $e_{ij}$. Note that vanishing $e$ must imply vanishing $e'$, for consistency.
Thus, there is no $e_{ij}$-independent term in $f_{ij}$.    Letting $E_{ij} \to
E_{ij} - \chi_{ij}$ and noting that to leading order
\be
e_{ij} = e'_{ij}- f_{ij} (\chi, e')\,,
\ee
 we have
\be
 S\big[E_{ij}-\chi_{ij} \, ,\;e'_{ij}+ \chi_{ij}-f_{ij}(\chi, e') \big] \ =
  S\big[E_{ij}\, ,\;e'_{ij}\big]\;.
\ee
Dropping the  primes, the condition of background independence to leading order in $ e$ becomes
\be
 S\big[E_{ij}-\chi_{ij} \, ,\;e_{ij}+ \chi_{ij}-f_{ij}(\chi, e) \big] \ =
  S\big[E_{ij}\, ,\;e_{ij}\big] +O(e^3)\;.
\ee
With the action calculated to cubic order, we can
only determine the leading terms in $f_{ij}$, those linear in $e$.
Our check of background independence  fixes $f_{ij}$ to be
\be
f_{ij} (\chi, e) =   \tfrac{1}{2}\,\bigl( \chi_{i}{}^{k}e_{kj}
+ \chi^{k}{}_j e_{ik}\bigr)+O(e^2)\;.
\ee
Recall that indices are raised with $G^{ij}$.

We now
explicitly check  the background independence to quadratic order
 by confirming that $\delta S=0$ up to terms cubic in fields  under the
variations
 \be
 \label{trans99}
  \delta e_{ij} = \chi_{ij}-\frac{1}{2}\bigl( \chi_{i}{}^{k}e_{kj}-
  \chi^{k}{}_j e_{ik}\bigr)\,,~~~
  \delta E_{ij} = -\chi_{ij}\,.
  \ee
  The background
  shift    changes   $G^{ij}$ by
  \be
  \delta G^{ij} \ = \ \chi^{(ij)} \ \equiv \ \frac{1}{2}\left(\chi^{ij}
  +\chi^{ji}\right)\;,
 \ee
 with $\chi^{ij}=G^{ik}\chi_{kl}G^{lj}$.
The full variation of $e_{ij}$ is needed for quadratic terms in the action, but only
the leading part is needed for the cubic terms.
For the computation it is useful to recall that
$\tilde{\partial}^{i}=\tfrac{1}{2}(\bar{D}^i-D^i)$, which implies that
 \bea\label{delD}
  \delta D_i = \frac{1}{2}\chi_{ik}(\bar{D}^k-D^k)\;, \qquad
  \delta\bar{D}_i = \frac{1}{2}\chi_{ki}(D^k-\bar{D}^k)\;.
 \eea
These lead to
 \bea\label{delbox}
  \delta\,\square \ \equiv \ \delta\left(G^{ij}D_i D_j\right) \ = \
  \chi^{ij}D_i \bar{D}_j\;.
 \eea
This identity in fact suffices to prove the background independence for
the  terms quadratic in the dilaton in the action (\ref{redef-action}).  Indeed, the following variations cancel
\begin{equation}
\begin{split}
\delta (-4d\,\square \,d) &=  -4 d\, \chi^{ij}D_i \bar{D}_j \,d\;,\\
\delta(4\,e_{ij} d\, D^i \bar D^jd   ) &= \, 4\,\chi_{ij} d\, D^i \bar D^jd \,.
\end{split}
\end{equation}
Checking the terms linear in $d$ is relatively straightforward. For structures quadratic in $e$, the variation of the bilinear terms gives
 \bea
  \delta {\cal L}^{(2)} \ = \ \frac{1}{4}\chi^{kl}e^{ij}D_k \bar{D}_l e_{ij}
  +\frac{1}{4}\chi^{mj}D_m e_{ij}\bar{D}_k e^{ik}+\frac{1}{4}\chi^{im}\bar{D}_m e_{ij}
  D_k e^{kj}\,.
 \eea
One can verify that, up to total derivatives, this variation is
cancelled by the variation of the terms cubic in $e_{ij}$. We note that these computations do
not require using the constraint $\partial_{i}\tilde{\partial}^{i}=0$.

Having checked background independence,
we can introduce
a  field ${\cal E}_{ij}$ that combines the background $E_{ij}$ and
the fluctuation $e_{ij}$ and is background independent in the sense that it is invariant under the transformations (\ref{trans99}).  We
find that
 \bea
 \label{eroier99}
{\cal E}_{ij} \equiv E_{ij} + e_{ij}+\frac{1}{2}e_i{}^{k}e_{kj} + {\cal O}(e^3)\,,
 \eea
has  this property.  Indeed,  under (\ref{trans99}),
\begin{equation}
\begin{split}
\delta {\cal E}_{ij} &=  \delta E_{ij} + \delta e_{ij}+\frac{1}{2}\delta
e_i{}^{k}e_{kj}
+\frac{1}{2}e_i{}^{k}\delta e_{kj} + {\cal O}(e^2)\\
&= -\chi_{ij} +\chi_{ij}-\frac{1}{2}\bigl( \chi_{i}{}^{k}e_{kj}-
  \chi^{k}{}_j e_{ik}\bigr) +\frac{1}{2}\chi_i{}^{k}e_{kj}
+\frac{1}{2}e_i{}^{k}\chi_{kj}  + {\cal O}(e^2)\\
&= 0+  {\cal O}(e^2)\,.
\end{split}
\end{equation}
Note that the index contraction $e_i{}^{k}e_{kj}$  on the right-hand
side of (\ref{eroier99}) breaks
the $O(D,D)$  covariance,
as will be discussed in subsection \ref{oddcovarianceproven}.  These are the first terms in the
full non-linear form of ${\cal E}$ in terms of
$E$ and $e$ that was found in  
\cite{Hull:2009zb,Michishita:2006dr} and which will be discussed in the next subsection.  It was shown
in \cite{Hull:2009zb} that upon setting all $\tilde \partial$
derivatives equal to zero, the field ${\cal E}$
has the familiar gauge
transformations associated with gravity and the antisymmetric tensor field.

\subsection{The gauge transformations in background independent form}\label{BindGaug}

In \cite{Hull:2009mi}, the action to cubic order in the fields was found to be (\ref{redef-action}), and the corresponding gauge transformations preserving this action were found to linear order in the fields. 
Reference~ \cite{Hull:2009zb} considered the situation in which the fields and gauge parameters are restricted to lie in a 
totally null subspace of 
the doubled space
i.e. a subspace in which all vectors are null and mutually orthogonal.  Such a restriction arises when the fields and parameters are all independent of $\tilde x$, or any configuration related to this by T-duality.  
The restriction guarantees that the strong form of the $\partial_M\partial^M=0$ constraint holds.
With the restriction,  the full  gauge transformations
to all orders in the fields are in fact quadratic in the fields~\cite{Hull:2009zb} and  given by:
\be
\label{finalgt0}
\begin{split}
\delta_\lambda
e_{ij}  &= ~   D_i \bar\lambda_j + \bar D_j \lambda_i \, \\
&~+   {1\over 2}\,(\lambda \cdot D + \bar \lambda \cdot \bar D) e_{ij}
+{1\over 2} \,   (D_i \lambda^k - \,  D^k \lambda_i)\, e_{kj}
    \, - e_{ik} \,\, {1\over 2}
   ( \bar D^k \bar \lambda_j - \bar D_j \bar \lambda^k )~~ \\[0.3ex]
&~ - {1\over 4}\, e_{ik}\,(D^l \bar \lambda^k +\bar D^k \lambda^l)\, e_{lj} \,,
\end{split}
\ee
where $ \lambda_i$ and $ \bar \lambda_i$ are independent real parameters.   
In \cite{Hull:2009zb} these gauge transformations were shown to close according to the
so-called C-bracket. To define this bracket we introduce
new gauge parameters  $\xi^i$ and $\tilde \xi_i$ defined by
 \bea\label{parashift}
  \lambda_{i}=-\tilde{\xi}_i+E_{ij}\xi^{j}\;, \qquad
  \bar{\lambda}_{i} = \tilde{\xi}_i +E_{ji}\xi^{j}\;.
 \eea
Using these parameters we can
write the gauge algebra in a manifestly $O(D,D)$ covariant form.
More precisely, if we assemble $\xi^i$ and $\tilde \xi_i$ into the fundamental $O(D,D)$ vector
\be
\xi^M  = \begin{pmatrix}  \tilde\xi_i \\ \xi^i \end{pmatrix}\,,
\ee
the closure of the gauge transformations defines the algebra
$\big[ \delta_{\xi_{1}},\delta_{\xi_2}\big] = -\delta_{[\xi_{1},\xi_2]_{\rm C}}$, where the C-bracket $[\cdot \,, \cdot ]_{\rm C}$ is given by
 \bea\label{Cbracket}
  \big[\xi_{1},\xi_{2}\big]^{M}_{\rm C}  
    \ = \ \xi^{N}_{1}\partial_{N}\xi^{M}_{2}
  -\frac{1}{2}\,\eta^{MN}\eta_{PQ}\,\xi^{P}_{1}\partial_{N}\xi^{Q}_{2}
  -\left(1\leftrightarrow 2\right)\;.
 \eea
The bracket (\ref{Cbracket}) was the starting point for the construction of Siegel \cite{Siegel:1993th}. In generalized geometry this is a bracket for a `Courant
bi-algebroid', and when restricted to fields and parameters independent of tilde coordinates, it reduces to the Courant bracket;
see the discussion in \cite{Hull:2009zb}. The background independent form of the gauge
transformations we are about to derive provide just a rewriting of the symmetry transformations, and therefore they will close according to the {\it same} bracket.  Thus the gauge algebra of the background independent theory can be said to be defined by the Courant bracket.

Let us now turn to the rewriting of the gauge transformations (\ref{finalgt0}).
We first write them as
   \be
\label{finalgtx}
\begin{split}  \delta_\lambda
e_{ij}  \ &= \  {1\over 2}\,(\lambda \cdot D + \bar \lambda \cdot \bar D) e_{ij}
     +\hat D_i \bar\lambda_j +{\hat { \bar D} }_j \lambda_i +{1\over 2} \,   (\hat D_i \lambda^k )\, e_{kj}
+ {1\over 2}
e_{ik} \,
    {\hat { \bar D} }_j \bar \lambda^k\,,
\end{split}
   \ee
   where
\be
 \hat D_i\equiv  D_i    - {1\over 2} e_{ik} \,
   \bar D^k, \qquad      {\hat { \bar D} }_j \equiv     {{ \bar D} }_j
  - {1\over 2} \,  e_{kj}     D^k\,.
\ee
The full  metric is
$g_{ij}=G_{ij}+h_{ij}$ and the
antisymmetric tensor gauge field is $b_{ij}=B_{ij}+ \hat{b}_{ij}$ where $G_{ij}$ and $B_{ij}$ are constant background fields. These are combined into
 \bea
 \label{eroier}
{\cal E}_{ij} \equiv E_{ij} + \e_{ij}\;,
 \eea
where $ \e_{ij}=h_{ij}+ \hat{b}_{ij}$. It was shown in  \cite{Hull:2009zb} that the field $ \e_{ij}$ is related to $ (e_{ij},d)$ by
  $ \e_{ij} =
  f_{ij}(e,d)$, where
\be
\label{mich1}
f =  \Bigl( 1- {1\over 2} \,e\Bigr)^{-1} e  \,,
\ee
so that
\be
\label{mich11}
\e =  F e  \,,
\ee
where we use matrix notation and
\be
\label{mich12}
F \equiv  \Bigl( 1- {1\over 2} \,e\Bigr)^{-1}   \,.
\ee
 Then the full non-linear form of (\ref{eroier99}) is
  \bea
 \label{eroier88}
{\cal E}_{ij}= E_{ij} + F_i{}^k (e) e_{kj}\;.
 \eea
Here and in the rest of this subsection, indices $i,j$ are raised and lowered using the background metric $G_{ij}$.

It is shown in appendix A that for any variation or derivative
\be
\delta {\cal E}= \delta \e= F\delta e F\;.
\ee
Rewriting the gauge transformations using this naturally
involves the derivatives
 \begin{equation}
 \begin{split}
  {\cal D}_{i}  & \ \equiv \ F_{ij}  \hat D_j\;, \\[0.5ex]
    \bar{{\cal D}}_i  & \ \equiv \  F_{ji} {\hat { \bar D} }_j\;.
      \end{split}
 \end{equation}
Remarkably
 these can be written (see appendix A) in manifestly background-independent form:
 \begin{equation}\label{groihfgruuclfss}
 \begin{split}
  {\cal D}_{i}  & \ = \ \partial_i-{\cal E}_{ik}\tilde{\partial}^k\;, \\[0.5ex]
    \bar{{\cal D}}_i  & \ = \ \partial_i+{\cal E}_{ki}\tilde{\partial}^k\;.
      \end{split}
 \end{equation}
From these derivatives one can reconstruct the ordinary partial derivatives through the formulae
  \bea\label{CalInv}
  \partial_{i}=\frac{1}{2}\left({\cal E}_{ji}{\cal D}^{j}+{\cal E}_{ij}\bar{\cal D}^{j}
  \right)\;, \qquad
  \tilde{\partial}^{i}=\frac{1}{2}\left(-{\cal D}^{i}+\bar{\cal D}^{i}\right)\;.
 \eea
It is useful to  rewrite the gauge transformations in terms of the gauge parameters $\xi^{i}$ and $\tilde{\xi}_i$ defined in (\ref{parashift}) \cite{Hull:2009zb}.
Then after some work (see appendix A)  the  gauge transformations take the manifestly background-independent form
 \bea
 \label{finalgt}  
 \delta {\cal E}_{ij} \ = \ {\cal D}_i\tilde{\xi}_{j}-\bar{{\cal D}}_{j}\tilde{\xi}_{i}
  +\xi^{M}\partial_{M}{\cal E}_{ij}
  +{\cal D}_{i}\xi^{k}{\cal E}_{kj}+\bar{\cal D}_{j}\xi^{k}{\cal E}_{ik}\;. 
 \eea
The gauge transformations can be expanded using
the derivatives $\partial_i,\tilde\partial^k$ as
 \bea
 \begin{split}
 \label{finalgt99}
\phantom{\Bigl(}~  \delta {\cal E}_{ij} \ &=~ \partial_i\tilde{\xi}_{j}-
\partial_{j}\tilde{\xi}_{i}
\\&~~
  +\tilde\xi_k \,\tilde\partial^k{\cal E}_{ij}
- \tilde\partial^k \tilde \xi_i\, {\cal E}_{kj} -
 \tilde\partial^k \tilde\xi_j\, {\cal E}_{ik} ~~~
  \\[0.5ex] &~~
  +\xi^k\partial_{k}{\cal E}_{ij}
 \, +\partial_{i}\xi^{k}\hskip1pt{\cal E}_{kj}+\partial_{j}\xi^{k}\,{\cal E}_{ik}
  \\&~~+ {\cal E}_{ik} \bigl( \tilde\partial^q \xi^k -  \tilde\partial^k \xi^q\bigr)
  {\cal E}_{qj}\;.   
  \end{split}   
  \eea
Along with the
standard Lie derivative
$ {\cal L}_{\xi}$
with respect to $\xi^{i}$,  we
introduce a dual Lie derivative with
respect to the tilde parameter $\tilde{\xi}_{i}$:
 \bea\label{dualLie}
  {\cal L}_{\tilde{\xi}}{\cal E}_{ij} \ \equiv
   \ \tilde{\xi}_{k}\tilde{\partial}^{k}
  {\cal E}_{ij}-\tilde{\partial}^{k}\tilde{\xi}_{i}\,{\cal E}_{kj}
  -\tilde{\partial}^{k}\tilde{\xi}_{j}\,{\cal E}_{ik}\; .
   \eea
The gauge transformations (\ref{finalgt99}) can
then
 be written more compactly as
 \bea\label{Liegauge}
  \delta_{\xi}{\cal E}_{ij} \ = \
   \partial_{i}\tilde{\xi}_{j}-\partial_{j}\tilde{\xi}_{i} + {\cal L}_{\xi}{\cal E}_{ij}
  +{\cal E}_{ik}
   \left(\tilde{\partial}^{q}\xi^{k}-\tilde{\partial}^{k}\xi^{q}\right){\cal E}_{qj}
    +{\cal L}_{\tilde{\xi}}{\cal E}_{ij}\;.
 \eea
For fields and parameters that are restricted to be independent of $\tilde x$ so that  $\tilde{\partial}^i=0$, this reduces to
 \bea
 \begin{split}
 \label{gaugag}
\phantom{\Bigl(}~  \delta {\cal E}_{ij} \ &={\cal L}_\xi {\cal E}_{ij}+  \partial_i\tilde{\xi}_{j}-
\partial_{j}\tilde{\xi}_{i}\;,
  \end{split} \eea
  which is the standard form of a diffeomorphism with the infinitesimal vector field
$\xi^{i}$  and a two-form gauge transformation with 
the infinitesimal one-form  
 $\tilde{\xi}_i$.
The gauge variation of the
inverse $ \tilde{{\cal E}}^{ij} \ \equiv \ \left({\cal E}^{-1}\right)_{ij}$
 is then
 \begin{equation}  
 \label{dualvar}
 \begin{split}
  \delta \tilde{\cal E}^{ij} &=
  -\tilde{\cal E}^{ik}\,\delta{\cal E}_{kl}\,\tilde{\cal E}^{lj} \\[0.5ex]
  &=~~ \tilde{\partial}^{i}\xi^{j}-\tilde{\partial}^{j}\xi^{i}
~ + \tilde{\xi}_{k}\tilde{\partial}^{k}\tilde{{\cal E}}^{ij}
  +\tilde{\partial}^{i}\tilde{\xi}_{k}\tilde{\cal E}^{kj}+\tilde{\partial}^{j}\tilde{\xi}_{k}
  \tilde{\cal E}^{ik}\\[1.0ex]
  &~-\tilde{\cal E}^{ik}\big(\partial_{k}\tilde{\xi}_{l}-\partial_{l}\tilde{\xi}_{k}\big)
  \tilde{\cal E}^{lj}+\xi^{k}\partial_{k}\tilde{\cal E}^{ij}-\tilde{\cal E}^{ik}\partial_{k}\xi^{j}
  -\tilde{\cal E}^{kj}\partial_{k}\xi^{i} \,.
  \end{split}
 \end{equation}
We can write this result as
 \begin{equation}\label{dualvar99}
  \delta \tilde{\cal E}^{ij} =
    \tilde{\partial}^{i}\xi^{j}-\tilde{\partial}^{j}\xi^{i}
  + {\cal L}_{\tilde{\xi}}\tilde{\cal E}^{ij}
  +
  \tilde{\cal E}^{ik}\big(\partial_{l}\tilde{\xi}_{k}-  \partial_{k}\tilde{\xi}_{l}\big)
  \tilde{\cal E}^{lj} +   {\cal L}_{\xi}\tilde{\cal E}^{ij}\;,
 \end{equation}
 where 
 \be
 \label{sdfafa}
  {\cal L}_{\tilde{\xi}}\tilde{\cal E}^{ij}= \tilde{\xi}_{k}\tilde{\partial}^{k}\tilde{{\cal E}}^{ij}
  +\tilde{\partial}^{i}\tilde{\xi}_{k}\tilde{\cal E}^{kj}+\tilde{\partial}^{j}\tilde{\xi}_{k}
  \tilde{\cal E}^{ik}\;.
  \ee 
By comparing with (\ref{Liegauge}) we infer that the role of $\xi$ and $\tilde{\xi}$ and of
${\cal E}$ and $\tilde{\cal E}$ are precisely interchanged; we shall see later that this interchange is a T-duality.
For fields and parameters that are restricted to be independent of $x$ so that  ${\partial}_i=0$, this reduces to
\be
 \delta \tilde{\cal E}^{ij} =
    \tilde{\partial}^{i}\xi^{j}-\tilde{\partial}^{j}\xi^{i}
  + {\cal L}_{\tilde{\xi}}\tilde{\cal E}^{ij}\;, 
  \ee  
  which is the standard form of a diffeomorphism of $\tilde x_i$ with the infinitesimal vector field
$\tilde \xi_{i}$  and a two-form gauge transformation with 
the infinitesimal one-form  
 ${\xi}^i$. (Note that here lower indices are contravariant and upper ones covariant, so that the signs in (\ref{sdfafa}) are the conventional ones.)

For the dilaton the full gauge transformation found in \cite{Hull:2009zb} is
\begin{equation}
\delta_\lambda  d~  = - {1\over 4}  (D\cdot \lambda +\bar D\cdot \bar \lambda )
+  {1\over 2}  (\lambda \cdot D+\bar\lambda \cdot \bar D) \,d\,,
\end{equation}
and can be rewritten as
\begin{equation}
\label{dilgag}
\delta_\lambda  d~  = - {1\over 2}  \partial_M \xi^M + \xi^M \partial_M \,d\,.\end{equation}

The gauge transformations are reducible -- there are ``symmetries of symmetries".
One can verify that parameters of the form
\begin{equation}
\xi^M = \partial^M \chi   ~~ \leftrightarrow~~  \tilde \xi_i = \partial_i \chi\,,
~~\xi^i = \tilde \partial^i \chi\,,
\end{equation}
generate  gauge transformations  (\ref{finalgt}), (\ref{dilgag}) that leave the fields unchanged, so that they
are trivial gauge transformations.
 The transport term is $\xi^{M}\partial_{M}{\cal E}_{ij}= \partial^M \chi  \partial_{M}{\cal E}_{ij}=0$ because of the strong form of the
 constraint.  For the rest of the terms
a small calculation is needed:
\begin{equation*}
\begin{split}
\delta {\cal E}_{ij} \ &= \ {\cal D}_i\partial_j \chi-\bar{{\cal D}}_{j}\partial_i \chi
  +{\cal D}_{i}\tilde\partial^k \chi {\cal E}_{kj}+\bar{\cal D}_{j}\tilde\partial^k \chi{\cal E}_{ik}\\
  &= \ -{\cal E}_{ik} \tilde \partial^k \partial_j \chi
  -{\cal E}_{kj} \tilde\partial^k\partial_i \chi
 +\partial_{i}\tilde\partial^k \chi {\cal E}_{kj}
  +\partial_{j}\tilde\partial^k \chi{\cal E}_{ik}
     \\
  &~~~~ - {\cal E}_{iq} (\tilde\partial^q \tilde\partial^k \chi ){\cal E}_{kj}
  +{\cal E}_{qj} (\tilde\partial^q\tilde\partial^k \chi){\cal E}_{ik}  = 0
\,.
  \end{split}
\end{equation*}

\subsection{Constructing the action}   \label{Constructingtheaction} %

The analysis in the previous subsection led to a background independent field
${\cal E}_{ij}$ defined in (\ref{eroier88}) (see also (\ref{eroier99})).  
In this section, we use this to construct a background
independent action that agrees with (\ref{redef-action}) to cubic order.
There is, however, a significant constraint.  The action
(\ref{redef-action}) only has
$O(D,D)$ consistent index contractions while
the redefinition (\ref{eroier88}) contains
$O(D,D)$ violating
contractions.  It is a stringent consistency test that the action
in terms of ${\cal E}_{ij}$ should only have
$O(D,D)$ consistent
contractions. We find a non-polynomical action that is background independent and has only consistent
index contractions.
In the following sections, we check that this action is duality covariant and gauge invariant, so that it must be the full non-linear action.

Given that the action to be constructed involves ${\cal E}$ it is natural
to modify the derivatives (\ref{groihfgruu8774}) and use the
calligraphic derivatives (\ref{groihfgruuclfss}) introduced in the previous subsection.
From these derivatives one can reconstruct the ordinary partial derivatives
through the formulae (\ref{CalInv}).
The constraint $\partial_M A \partial^M B = 0$, or equivalently
$\partial_i A \tilde \partial^i B + \tilde\partial^i A  \partial_i B=0$ takes a simple form using
calligraphic derivatives.   A short calculation shows that it is equivalent
to
\be
\label{condflk}
{\cal D}^i  \hskip-2pt A \, {\cal D}_i  B  = \bar {\cal D}^i  \hskip-2pt A
\, \bar {\cal D}_i  B \,.
\ee
 We also must have for any $A$:
\be
\partial_i \tilde \partial^i A =0
\,,
\ee
 or equivalently
$
\partial_M  \partial^M A = 0$.
Using (\ref{CalInv})
this constraint can be written with
calligraphic derivatives~as
\be
{\cal E}_{ij} \left(\bar{\cal D}^{j}\bar{\cal D}^{i}
-{\cal D}^{i}{\cal D}^{j}
-\bar{\cal D}^{j}{\cal D}^{i}+
{\cal D}^{i}\bar{\cal D}^{j}
\right) A=0\;.
\ee
We define $g^{ij}$ as the inverse of the metric $g_{ij} = {1\over 2} ({\cal E}_{ij} + {\cal E}_{ji})$.  A short calculation shows that
 \bea
g^{ij} =   G^{ij}  - e^{(ij)} + {1\over 4} e^{ik} e^j_{~k} +{1\over 4}
 e^{ki}e_{k}{}^{j}+ {\cal O} (e^3)\;.
 \eea
Indices on the fluctuation field $e_{ij}$ are raised
with the constant background (inverse) metric $G^{ij}$.
The calligraphic derivatives can be written in terms of
$D, \bar D$ derivatives and fluctuations \begin{equation}
 \begin{split}
  {\cal D}_{i}  & ~ = ~
  D_i  - {1\over 2} e_{ik}  (\bar D^k - D^k)
  -\frac{1}{4}e_{i}{}^{k}e_{kl}(\bar{D}^{l}-D^{l})+ {\cal O} (e^3)\;, \\[0.5ex]
    \bar{{\cal D}}_i  & ~ =  ~ \bar D_i  + {1\over 2} e_{ki}  (\bar D^k - D^k)
   +\frac{1}{4}e_{k}{}^{l}e_{li}(\bar{D}^{k}-D^{k})+ {\cal O} (e^3)\;.
  \end{split}
  \end{equation}
The indices on the calligraphic derivatives will be raised with the
full metric $g^{ij}$.  We thus find:
 \begin{equation}
\begin{split}
  {\cal D}^{i}  &\equiv  g^{ij} {\cal D}_j=
  D^i  - {1\over 2} e^{ij}\,\bar D_j   - {1\over 2} \,e^{ji} D_j+ {\cal O} (e^2)\;,
  \\[0.5ex]
    \bar{{\cal D}}^i &\equiv  g^{ij} \bar{\cal D}_j= \bar D^i  - {1\over 2} e^{ki} D_k - {1\over 2}  e^{ik} \bar D_k+ {\cal O} (e^2)\;.
    \end{split}
 \end{equation}

An action written only using the calligraphic derivatives, $g^{ij}$,
${\cal E}_{ij}$ and $d$ will be manifestly background independent.
As discussed in the following subsection, it will be duality covariant provided
the index contractions all satisfy the rules of contraction formulated in~\cite{Hull:2009mi}.
  Our strategy now is to seek such a background independent duality covariant action that agrees
  with the action (\ref{redef-action}) to cubic order.
The dilaton theorem states that a constant shift of the dilaton
is equivalent to a change of the coupling constant. This theorem is manifest in
actions where the dilaton appears in an exponential prefactor that multiplies
all terms in the Lagrangian, and all other occurrances of the dilaton
involve its derivatives.
We thus aim for an action where the  overall multiplicative factor
takes the conventional form $e^{-2d}$ and elsewhere the dilaton
appears with derivatives.

We now begin the computation.
The $(-4d\,\square d)$ term in the action (\ref{redef-action}) could come from a term $4 e^{-2d}{g}^{ij}{\cal D}_{i}d\,{\cal D}_{j}d$.  Expansion in fluctuations gives the desired quadratic term and extra cubic terms:
 \be
  4 e^{-2d}g^{ij}{\cal D}_{i}d\,{\cal D}_{j}d =
  -4d\,\square d+4e^{ij}d D_i\bar{D}_j d+4d^2\square d
 -2d^2D^{i}\bar{D}^{j}e_{ij}  + (\hbox{td})\;,
\ee
where (td) stands for
$\partial _i$ and  $\tilde \partial ^i$ total derivatives,
$(\hbox{td})=\partial _M v^M$
for some $v^M$.
 These terms can be ignored as we integrate
to form the action.  For the quadratic terms mixing the dilaton and $e_{ij}$ we use
\bea
e^{-2d} \bigl(
 {\cal D}^{i}d \,\bar{\cal D}^{j}{\cal E}_{ij}
  +\bar{{\cal D}}^{j}d\,{\cal D}^{i}{\cal E}_{ij} \bigr) &=&
  -2d D^i \bar{D}^j e_{ij} + d e^{ij}(D_iD^k e_{kj}
  +\bar{D}_j\bar{D}^k e_{ik})\nonumber \\
  &&+\frac{1}{2}d\left(D^{i}e_{ij}\right)^2+\frac{1}{2}d
  \left(\bar{D}^{j}e_{ij}\right)^2\\
  \nonumber
  &&+\frac{1}{2}d\bar{D}^{j}e^{il}\,\bar{D}_{l}e_{ij}+\frac{1}{2}d\,D^{i}e^{kj}\,
  D_{k}e_{ij}+2d^2\,D^{i}\bar{D}^{j}e_{ij}+ (\hbox{td})\;. \nonumber
\eea
To obtain the terms quadratic in $e_{ij}$ we need three structures
 \bea
  \frac{1}{4}e^{-2d}g^{kl}{\cal D}^{j}{\cal E}_{ik}{\cal D}^{i}{\cal E}_{jl}
  &=& \frac{1}{4}(D^{i}e_{ik})^2-\frac{1}{4}e_{ij}\,D^{k}e^{il}\,\bar{D}^{j}e_{kl}
  -\frac{1}{2}d\,D^{j}e_{ik}\,D^{i}e_{j}{}^{k}+ (\hbox{td})\;,\\
  \frac{1}{4}e^{-2d}g^{kl}\bar{\cal D}^j{\cal E}_{ki}\bar{\cal D}^{i}{\cal E}_{lj}
  &=& \frac{1}{4}(\bar{D}^{i}e_{ki})^2-\frac{1}{4}e_{ij}D^ie_{kl}\bar{D}^{l}e^{kj}
  -\frac{1}{2}d\,\bar{D}^{j}e^{ki}\,\bar{D}_{i}e_{kj}+ (\hbox{td})\;, \\
  -\frac{1}{4}e^{-2d}\,
   g^{ik}g^{jl} \,
    {\cal D}^p{\cal E}_{kl}\,{\cal D}_{p} {\cal E}_{ij}
  &=& \frac{1}{4}e^{ij}\square e_{ij}+\frac{1}{4}e^{ij}D_{i}e^{kl}\,
  \bar{D}_{j}e_{kl} \\ \nonumber
  &&+\frac{1}{4}d\,D^{i}e^{jk}\,D_{i}e_{jk}+\frac{1}{4}d\,\bar{D}^{i}e^{jk}\bar{D}_{i}
  e_{jk}+ (\hbox{td})\;.
 \eea
In all of the above equations we are ignoring terms of higher order  than cubic in the fields. We also note that 
here  the constraint
$\partial_M\partial^M=0$ 
 is not used. The five structures above, added together,
give the action
\bea
\label{THEAction}
 \begin{split}\hskip-10pt
  S \ = \ \int \,dx d\tilde{x}~
  e^{-2d}\Big[&
  -\frac{1}{4} \,g^{ik}g^{jl}   \,   {\cal D}^{p}{\cal E}_{kl}\,
  {\cal D}_{p}{\cal E}_{ij}
  +\frac{1}{4}g^{kl} \bigl( {\cal D}^{j}{\cal E}_{ik}
  {\cal D}^{i}{\cal E}_{jl}  + \bar{\cal D}^{j}{\cal E}_{ki}\,
  \bar{\cal D}^{i}{\cal E}_{lj} \bigr)~
\\ &    + \bigl( {\cal D}^{i}\hskip-1.5pt d~\bar{\cal D}^{j}{\cal E}_{ij}
 +\bar{{\cal D}}^{i}\hskip-1.5pt d~{\cal D}^{j}{\cal E}_{ji}\bigr)
 +4{\cal D}^{i}\hskip-1.5pt d \,{\cal D}_{i}d ~\Big]\;,
 \end{split}
 \eea
and, by construction, 
reproduce all quadratic terms in the action (\ref{redef-action}).
Remarkably, they also reproduce
precisely the cubic terms in (\ref{redef-action}), with no further terms needed!

Terms have been grouped in parentheses  to make the $\mathbb{Z}_2$
symmetry of the action manifest.  This symmetry, discussed in~\cite{Hull:2009mi},
 exchanges the
indices in ${\cal E}$, exchanges barred and unbarred derivatives,
and leaves the dilaton invariant. For the first and last term the constraint
(\ref{condflk}) is needed to guarantee the $\mathbb{Z}_2$
symmetry of the action.

Making all metrics explicit, the action is
\bea
\label{THEActionMET}
\begin{split}
  S \ = \ \int \,dx d\tilde{x}~
  e^{-2d}\Big[&
  -\frac{1}{4} \,g^{ik}g^{jl}  \,g^{pq} \, \Bigl(
  {\cal D}_{p}{\cal E}_{kl}\,  {\cal D}_{q}{\cal E}_{ij}
  -  {\cal D}_{i}{\cal E}_{lp} {\cal D}_{j}{\cal E}_{kq}
  - \bar{\cal D}_{i}{\cal E}_{pl}\,  \bar{\cal D}_{j}{\cal E}_{qk}
  \Bigr)~
\\ &    + g^{ik}g^{jl}\bigl( {\cal D}_{i} d~
\bar{\cal D}_{j}{\cal E}_{kl}
 +\bar{{\cal D}}_{i} d~{\cal D}_{j}{\cal E}_{lk}\bigr)
 +4g^{ij}{\cal D}_{i}d \,{\cal D}_{i}d ~\Big]\;.\phantom{\biggl(}
 \end{split} 
 \eea
This action is background-independent and, as we will show in the next subsection, it is invariant under T-duality.
Any non-trivial background independent term with two derivatives that could be added to $S$ would also contribute to the quadratic and cubic actions and so would spoil the agreement with (\ref{redef-action}).
Thus we 
conclude  that $S$ 
is the complete background independent action! Indeed, we will
establish that this is the case, and in particular show that $S$ is invariant under the gauge transformations (\ref{finalgt}), is duality invariant, and reduces to the standard action for suitably restricted fields.

\subsection{O(D,D) invariance}  \label{oddcovarianceproven}

The action (\ref{redef-action}) was proven in~\cite{Hull:2009mi} to
possess T-duality covariance.   In that work a notation was used that
allowed to deal with T-duality transformations in a theory with
both compact and non-compact directions.
The spacetime has dimension
$D= n+ d$
and is
the product of $n$-dimensional Minkowski space
$\mathbb{R}^{n-1,1}$ and a torus $T^d$.  Although we write
$O(D,D)$ matrices,  the ones that are used describe T-dualities that belong to the
$O(d,d)$ subgroup associated with the torus.
The subgroup preserving the periodicity conditions of the doubled torus is the discrete group $O(d,d;\Z)$, and this is the proper T-duality group.
It is interesting to note, however, that our action could also be used for a situation in which all coordinates are non-compact but are nonetheless doubled, so that we have $2D$ non-compact coordinates $x, \tilde x$. For such a set-up, the theory formally has a continuous $O(D,D)$ symmetry and for this reason we will not always distinguish here between the discrete and continuous groups, and will refer to both as T-dualities.
We stress, however,  that the
 case relevant to string theory
is that in which  there are $d$ compact coordinates and their doubles,
with T-duality group $O(d,d;\Z)$.

The $O(D,D)$ matrices $h$ take the form
\be
\label{odddefstr}
 h= \begin{pmatrix} a& b \\ c& d \end{pmatrix}
\, ,   ~~~ h^{-1}= \begin{pmatrix} d^t& b^t \\ c^t& a^t \end{pmatrix}, ~~
\,\, ~~   h^t \eta h = \eta \,,   ~~\eta= \begin{pmatrix} 0& I \\ I& 0 \end{pmatrix}\,.
\ee
If the background field $E$ is 
viewed as a parameter in the action (\ref{redef-action}),  
then any change of its value  would represent,
a priori, a different theory. 
The $T$-duality covariance of (\ref{redef-action}) established the equivalence of two such actions written with different backgrounds
$E$ and $E'$ related by T-duality. Of course, if we transform both the
background and the fluctuations, we get an invariance of the
action.  In a background independent
formulation, like that of (\ref{THEAction}), the 
 invariance is natural  since we transform the full fields that include background and fluctuations.
The doubled coordinates 
$X^M$
transform linearly under $O(D,D)$
\be
 \label{vecx}
{X'}\equiv 
 \begin{pmatrix} \,\tilde x'\, \\[0.6ex] {x'} \end{pmatrix}
= h X =   \begin{pmatrix} a & b \\ c & d \end{pmatrix} \begin{pmatrix} \,\tilde x
\, \\[0.6ex] x
  \end{pmatrix}\,.
\ee
The derivatives $ \partial _M=(\tilde{\partial}^{i},\partial_{i})$
then transform as
\be
\partial ' = (h^{-1})^t \,\partial  \, .  
\ee
It was argued  in~\cite{Hull:2009mi} that
the dilaton should be 
an $O(D,D)$  \lq scalar':
\be
 \label{dscalar}
d'(X')=d(X)\,,
\ee
while
the transformation of ${\cal E}_{ij}$  should take the fractional linear form
\be
\label{eeprime}
{\cal E}' (X')
=   ( a {\cal E}(X) + b ) (c {\cal E}(X) + d)^{-1} \,,
\ee
when written in terms of
$D\times D$ matrices. For the special case in which the  fields are independent of the toroidal coordinates and their duals, these are the familiar Buscher transformations.

It is useful to introduce matrices $M(X)$ and $\bar M(X)$ that control the transformations
of certain tensors:
\be
\label{mbarm}
M(X)\equiv  d^t - \,{\cal E}(X) \, c^t\, ,~~
\bar M(X) \equiv  d^t  + {\cal E}^t(X) c^t  \,.
\ee
Similar matrices were defined in \cite{Hull:2009mi} but with the constant
background $E$ instead of the position-dependent ${\cal E}(X)$.
We record some useful identities:
\be
\label{jnbvgn}
\begin{split}
b^t - {\cal E} a^t &= -  M (X) {\cal E}'\,,\\[1.0ex]
b^t + {\cal E}^t a^t &= ~\bar M(X)\, {\cal E}'^t \,.
\end{split}
\ee

The $O(D,D)$ transformations of the inverse metric $g^{-1}$ can be given in two equivalent forms, one in terms of $M$ and one in terms of $\bar M$.
The transformations are found following the arguments
used in~\cite{Kugo:1992md} and give
\be
\label{fsetmat4678599}
g^{-1} = (\bar M^t )^{-1}\, g^{\prime -1}  \,\bar M^{-1}\,,~~~
g^{-1} = (M^t)^{-1}  \, g^{\prime -1}\, M^{-1} \,.
\ee
Here $M$ and $\bar M$ depend on $X$, $g$ depends on $X$ and $g'$ on $X'$, so that
e.g. the second expression is
$g^{-1} (X)= (M^t)^{-1}(X)  \, g^{\prime -1}(X')\, M^{-1}(X) $.
The matrices $M$ and $\bar M$
also
control the transformation
of the calligraphic derivatives:
\be
\label{vmswp}
{\cal D}_i  =  M_{i} {}^k \,{\cal D}'_k \,,  ~~~~
\bar{\cal D}_i  =  \bar M_{i}{}^k \, \bar{\cal D}'_k \,.
\ee

We will refer to quantities transforming in this way with
the matrices $M$ and $\bar M$
acting on the indices
as transforming covariantly or tensorially under $O(D,D)$.
Our definition may be a little perverse, as it differs from the usual usage for tensor representations of $O(D,D)$ in which transformations such as (\ref{vecx}) would be termed covariant.  Then
 the inverse metric and the calligraphic derivatives transform tensorially.
The transformation properties can be
indicated
 using barred and unbarred indices, as in~\cite{Hull:2009mi}. In this notation, we use lower unbarred indices for indices transforming with $M$,
 lower barred indices for indices transforming with $\bar M$,
  upper unbarred indices for indices transforming with $M^{-1}$ and
  upper barred indices for indices transforming with $\bar M^{-1}$.

In this terminology, an $O(D,D)$ tensor
$T_{i_1 \ldots i_p, \bar j_1 \ldots \bar j_q} (X)$ has a number of unbarred and barred indices and
transforms as follows:
\be
\label{oddtensor}
T_{i_1 \ldots i_p, \bar j_1 \ldots \bar j_q} (X) =  M_{i_1}^{~ k_1} \ldots M_{i_p}^{~ k_p}
\,\bar M_{\bar j_1}^{ ~ \bar l_1} \ldots \bar M_{\bar j_q}^{~ \bar l_q}\, \, {T^\prime}_{k_1 \ldots k_p, \bar l_1 \ldots \bar l_q} (X')\,.
\ee
Here and throughout this section, $M=M(X)$ and $\bar M=\bar M(X)$ depend on $X$, not $X'$.
In other sections of the paper, we drop the bars and just write $T_{i_1 \ldots i_p, j_1 \ldots j_q} (X)$.
An $O(D,D)$ tensor
with upper indices  transforms as
\be
\label{oddtensorupper}
U^{i_1 \ldots i_p, \bar j_1 \ldots \bar j_q} (X) =   \,{ U^\prime}^{k_1 \ldots k_p, \bar l_1 \ldots \bar l_q} (X')\,\,(M^{-1})_{k_1}^{~ i_1} \ldots (M^{-1})_{k_p}^{~ i_p}
\,(\bar M^{-1})_{\bar l_1}^{ ~ \bar j_1} \ldots (\bar M^{-1})_{\bar l_q}^{~ \bar j_q}\,.
\ee
With index notation and explicit barred and unbarred indices, the transformation of $g^{-1}$
in (\ref{fsetmat4678599}) is:
\be
\label{fsetmatvmvgn99}
g^{\bar i \,\bar j} =   (\bar M^{t^{-1}})^{\bar i}_{~\bar p}\,
g^{\prime\,\bar p \bar q}  \,(\bar M^{-1})_{\bar q}^{~\bar j}\,,~~~
g^{ij} = ( M^{t^{-1}})^i_{~p} \, g^{\prime \,pq}\, (M^{-1})_q^{~j} \,,
\ee
or eliminating the transposes,
\be
\label{fsetmatvmvgn9998}
g^{\bar i\,\bar j} =   (\bar M^{-1})^{~\bar i}_{\bar p}\,
g^{\prime\, \bar p\bar q}  \,(\bar M^{-1})_{\bar q}^{~\bar j}\,,~~~
g^{ij} = ( M^{-1})^{~i}_{p} \, g^{\prime \,pq}\, (M^{-1})_q^{~j} \,.
\ee
Note that the inverse metric can be viewed as an object with two
unbarred upper indices or two barred upper indices. This means that
$g^{\bar i\,\bar j} = g^{ i j}$ when $\bar i = i$ and $\bar j = j$.
The transformations (\ref{oddtensor}) and
(\ref{oddtensorupper}) are then  consistent with using the metric $g$ to raise and lower indices.
Our calligraphic derivatives transform as (\ref{vmswp}),
so the index on the $\bar{\cal D}$ is of barred type 
and the index on ${\cal D}$ is unbarred.

The field ${\cal E}(X)$ transforms projectively, as indicated in (\ref{eeprime}), but
variations of the field transform as an $O(D,D)$ tensor.
Defining ${\cal E}'+\delta {\cal E}'$ as the image of
${\cal E}+\delta {\cal E}$ under an $O(D,D)$  transformation, one
finds that $\delta {\cal E}_{ij}(X)$ and $\delta {\cal E}'_{ij} (X')$ are related by
\be
\label{pssiu}
\delta {\cal E}(X)=M(X) \,\delta {\cal E}'(X') \bar M^t (X)\,,
\ee
which means that the first index in
$\delta{\cal E}$
is unbarred and the second is barred.
This relation applies to any variation or  derivative:
\be
\partial_i{\cal E}=M \,\partial_i {\cal E}' \bar M^t\,,
~~~\tilde\partial^i{\cal E}=M \,\tilde\partial^i {\cal E}' \bar M^t\,,
\ee
and therefore it also applies for calligraphic derivatives
in the form
\be
{\cal D}_i{\cal E}=M \,{\cal D}_i {\cal E}' \bar M^t\,,
~~~\bar{\cal D}_i{\cal E}=M \,\bar{\cal D}_i {\cal E}' \bar M^t\,.
\ee
Note that the ${\cal D}$ and $\bar{\cal D}$ derivatives are identical on both sides of these
transformations; the ${\cal E}$'s within the derivatives have not been
transformed.  
Inserting all  indices in the above and using the transformation (\ref{vmswp})
of the calligraphic derivatives
\be
{\cal D}_i{\cal E}_{jk}= M_{i}^{~q}\,M_{j}^{~p}\, \bar M_{k}^{~\ell} \,\,{\cal D}'_q {\cal E}'_{p\ell} \,,
~~~\bar{\cal D}_i{\cal E}_{jk}=\bar M_{i}^{~q}M_{j}^{~p}\,
\bar M_{k}^{~\ell}\bar{\cal D}_q^{\prime} {\cal E}'_{p\ell} \,.
\ee
Thus calligraphic derivatives of ${\cal E}$ are $O(D,D)$ tensors!
With barred and unbarred indices, these are ${\cal D}_i{\cal E}_{j \bar k}$ and $\bar{\cal D}_{\bar i}{\cal E}_{j \bar k}$.

Using barred and unbarred indices, the Lagrangian in
 (\ref{THEAction})
 can be written as
 \bea
\label{THELag}
 \begin{split}\hskip-10pt
  {\cal L} \ =~
  e^{-2d}\Big[&
  -\frac{1}{4} \,g^{ik}g^{\bar j \bar l}   \,   {\cal D}^{p}{\cal E}_{k\bar l}\,
  {\cal D}_{p}{\cal E}_{i\bar j}
  +\frac{1}{4} \bigl(g^{\bar k \bar l} {\cal D}^{j}{\cal E}_{i \bar k}
  {\cal D}^{i}{\cal E}_{j \bar l}  +
  g^{kl} \bar{\cal D}^{\bar j}{\cal E}_{k \bar i}\,
  \bar{\cal D}^{\bar i}{\cal E}_{l \bar j} \bigr)~
\\[0.5ex] &    + \bigl( {\cal D}^{i}\hskip-1.5pt d~\bar{\cal D}^{\bar j}{\cal E}_{i \bar j}
 +\bar{{\cal D}}^{\bar i}\hskip-1.5pt d~{\cal D}^{j}{\cal E}_{j \bar i}\bigr)
 +4{\cal D}^{i}\hskip-1.5pt d \,{\cal D}_{i}d ~\Big]\; .
 \end{split}
 \eea
We  see that all appearances of ${\cal E}$ are with
single calligraphic derivatives, thus as $O(D,D)$ tensors.
More than one derivative or no derivatives on ${\cal E}_{ij}$
would imply complications with $O(D,D)$ covariance.
Since all  contractions in the above Lagrangian  are between upper and lower unbarred indices or
 between upper and lower barred indices,
 the Lagrangian is an $O(D,D)$ scalar:
 \be \label{lscalar} 
  {\cal L '}(X ')=  {\cal L}(X)\,.
  \ee
  As the measure is invariant, this establishes the $O(D,D)$ invariance
  of the action.

\medskip
The $O(D,D)$ covariance of the gauge transformations
is more nontrivial to verify.  We have (\ref{finalgt}) that reads
 \bea\label{gaugetr99}
  \delta {\cal E}_{ij}
  =
  {\cal D}_i\tilde{\xi}_{j}-\bar{{\cal D}}_{j}\tilde{\xi}_{i}
  +{\cal D}_{i}\xi^{k}{\cal E}_{kj}+\bar{\cal D}_{j}\xi^{k}{\cal E}_{ik}
  +\xi^M  \partial_M {\cal E}_{ij}\;.
 \eea
Here the gauge parameters can be grouped into the $O(D,D)$ `vector'
\be\label{xiM}
\xi^M  = \begin{pmatrix}  \tilde\xi_i \\ \xi^i \end{pmatrix}\,.
\ee
We have the transformation  $\xi' = h\, \xi$, which   implies
that $\xi = h^{-1} \xi'$.  Using (\ref{odddefstr}),   we find 
for the components in (\ref{xiM}) 
\be
\label{gfti}
\tilde \xi  =  d^t  \,\tilde \xi'   +  b^t  \,\xi'  \,, ~~~~ \xi =
c^t \,\tilde \xi' + a^t  \, \xi'\,.
\ee
Using this, (\ref{vmswp}), (\ref{pssiu}), and noting that the operator  $\xi^M \partial_M$ is
$O(D,D)$ invariant,
 (\ref{gaugetr99}) becomes
 \bea\label{gaugetr9vmsprt}
 \begin{split}
M_{i}^{~k}  \bar M_j^{~\ell }\,\delta {\cal E}'_{k\ell}  ~
  =&
~~~~  M_i^{~k}{\cal D}'_k (d^t  \,\tilde \xi'   +  b^t  \,\xi' )_{j}
- \bar M_j^{~\ell }\bar{{\cal D}}'_{\ell}
 (d^t  \,\tilde \xi'   +  b^t  \,\xi' )_{i}  \\
&  +M_i^{~k}{\cal D}'_k(c^t \,\tilde \xi' + a^t  \, \xi'  )^{p}{\cal E}_{pj}
+\bar M_j^{~\ell }\bar{\cal D}'_{\ell}(c^t \,\tilde \xi' + a^t  \, \xi'  )^{k}{\cal E}_{ik}
 \\
 &  +M_i^{~k} \bar M_j^{~\ell }({\xi'}^M  \partial'_M {\cal E}'_{k\ell} ) \;.
\end{split}
 \eea
Expanding out the derivatives and combining terms we readily find
 \bea\label{gaugetr9vmsvprtty}
 \begin{split}
 M_{i}^{~k}  \bar M_j^{~\ell }\,\delta {\cal E}'_{k\ell}  ~
  =&
~~~~   M_{i}^{~k} \,{\cal D}'_k\tilde \xi'_\ell\,(d + c{\cal E})_{\ell j}
-
  \,\bar M_j^{~\ell }
  \bar{{\cal D}}'_{\ell}\tilde \xi'_k \, (d^t  - {\cal E} c^t) _{ik}
  \\
&  + M_{i}^{~k} \, {\cal D}'_k\xi'^\ell  (a {\cal E} + b)_{\ell j}
+  \,\bar M_j^{~\ell }\bar{\cal D}'_{\ell} \xi^{'k} ({\cal E} a^t- b)_{ik}  \\
 &  + M_{i}^{~k} \bar M_j^{~\ell }({\xi'}^M  \partial'_M {\cal E}'_{k\ell} ) \;.
\end{split}
 \eea
We now identify using (\ref{mbarm}) and (\ref{jnbvgn})
 \bea\label{gaugetr9vmsvprttyps}
 \begin{split}
 M_{i}^{~k}  \bar M_j^{~\ell }\,\delta {\cal E}'_{k\ell}  ~
  =&
~~~~  M_{i}^{~k}\bar M_j^{~\ell } ( \,{\cal D}'_k\tilde \xi'_\ell -
\,\bar{{\cal D}}'_{\ell}\tilde \xi'_k )
  \\
&  + M_{i}^{~k} \, {\cal D}'_k\xi'^p  ({\cal E}' \bar M^t)_{p j}
+ \bar M_j^{~\ell } (M{\cal E}')_{ip}  \,\bar{\cal D}'_{\ell} \xi^{'p}  \\
 &  + M_{i}^{~k} \bar M_j^{~\ell }({\xi'}^M  \partial'_M {\cal E}'_{k\ell} ) \,.
\end{split}
 \eea
This finally yields
 \bea\label{gaugetr9vmsvprttypsy}
 \begin{split}
 M_{i}^{~k}  \bar M_j^{~\ell }\,\delta {\cal E}'_{k\ell}
 = M_{i}^{~k}  \bar M_j^{~\ell }\,( \,{\cal D}'_k\tilde \xi'_\ell\, -
 \,\bar{{\cal D}}'_{\ell}\tilde \xi'_k  +
  {\cal D}'_k\xi'^p  {\cal E}'_{p\ell}
  +\,\bar{\cal D}'_{\ell} \xi^{'p} {\cal E}'_{kp}
  +{\xi'}^M  \partial'_M {\cal E}'_{k\ell}   )\,.
\end{split}
 \eea
Deleting the matrices $M$ and $\bar M$ we find that the
gauge transformations for the primed variables take exactly the same form
as those for the unprimed variables. This confirms the
$O(D,D)$  covariance of the gauge transformations.  Thus the
action and the gauge transformations are consistent with $O(D,D)$
symmetry.

\subsection{The strong constraint and restricted fields}\label{tscarf}

The strong constraint
 requires that all fields
and gauge parameters and all of their possible products and powers
are annihilated by $ \partial^{M}\partial_{M}$.  We will show here that
this constraint implies that all fields and gauge parameters are restricted in the sense of~\cite{Hull:2009zb}: they
 depend only on the coordinates of  a totally null subspace $N$, so that the theory is related by an $O(D,D)$ transformation to one in which
all fields and gauge parameters depend on $x$ but do not depend on $\tilde x$.
Note that here we only consider classical field theory, so that all products of fields are conventional classical products.

The constraint means  that for any two fields $A$ and $B$,
 $ \partial^{M}\partial_{M}A = \partial^{M}\partial_{M} B=0$
as well as  $ \partial^{M}\partial_{M} (AB)=0$, which requires
 \bea\label{strongconstr}
  \partial^{M}A\,\partial_{M}B \ = \ 0\;,
 \eea
or equivalently
$ \partial_{i}A\,\tilde{\partial}^{i}B+\partial_{i}B\,\tilde{\partial}^{i}A = 0.$
Indeed, if this is true, $ \partial^{M}\partial_{M}$ will annihilate all multiple products or powers of fields.

 Consider first a field
comprised of a single Fourier mode
 \bea
  A(\tilde{x}_{i},x^{i}) \ =
  A\,e^{i(\tilde p^{i}\tilde{x}_{i}+p_{i}x^{i})}\;.
 \eea
As $(\tilde{x},x)$ transforms as a vector under $O(D,D)$, the `dual vector'
 \bea\label{fouriervec}
 P_M \ \equiv \ \begin{pmatrix} \,\tilde p^i \, \\[0.6ex] p_i  \end{pmatrix}
 \eea
also transforms as a vector under $O(D,D)$.
In terms of (\ref{fouriervec}), the constraint $\partial^M\partial_M A=0$ implies
 \bea
  \tilde{\partial}^{i}\partial_{i}A \ = \ 0 \qquad \Leftrightarrow \qquad
  \tilde p^{i}p_{i} \ = \ 0
   \qquad \Leftrightarrow \qquad
  P\cdot P\ \equiv \ \eta^{MN}P_{M}P_{N} \ = \ 0\;.
 \eea
The momentum vector $P$ corresponding to each Fourier component of each field is then a null vector.
Moreover, the constraint (\ref{strongconstr}) implies that any two momentum vectors $P_{\alpha}, P_{\beta} $
 associated with two
Fourier components (of the same or different fields)  must be orthogonal, so that
\bea\label{nulls}
   P_{\alpha}\cdot P_{\beta} \ = \ 0 \; \qquad \forall ~ \alpha, \beta\;.
 \eea
Then all momenta $P_\alpha$ must lie in a subspace of $\mathbb{R}^{2D}$ that is totally null or isotropic, i.e. any two vectors in the space are both null and are mutually orthogonal.

The maximal dimension for such  an isotropic subspace is $D$
and a maximal isotropic subspace is one of dimension $D$ -- it is maximal as there can be no isotropic space of dimension larger than $D$.
The canonical example of a maximal isotropic subspace is the subspace $T^*\subset \R^{2D}$
spanned by $P$'s with  $\tilde p=0$, so that
\bea\label{fouriervecpo}
 P_M \ \equiv \ \begin{pmatrix} \,0 \, \\[0.6ex] p_i  \end{pmatrix}\;.
 \eea
 This is the momentum space when
 all fields depend on $x$ and not on $\tilde x$, so that all momenta lie in
 $T^*$.
Any isotropic subspace is a subspace of a
maximal one.\footnote{Suppose $E$ is an isotropic subspace of dimension $n<D$.
Then it is straightforward to construct a maximal isotropic subspace $L$ containing $E$.
Let $E_\perp$ be the space of vectors orthogonal to $E$; this contains $E$ so can be written as $E_\perp=E\oplus F$ for some  space $F$ which has dimension $2(D-n)$.
Then it is straightforward to show that $K=T^*\cap F$ is an isotropic subspace of dimension $D-n$ and that $L=E\oplus K$ is  isotropic and of dimension $D$ and so is a
 maximal isotropic subspace
 containing $E$. Other such maximal isotropic subspaces  can be obtained by acting on $L$ with $O(D,D)$ transformations preserving $E$.}
Then the strong  constraint implies all momenta must lie in some maximal isotropic subspace $N$  
 of dimension $D$.
  This implies that all the fields and  gauge parameters are restricted to depend only on the coordinates of a totally null $D$-dimensional subspace 
  of the $2D$-dimensional  double space.
In~\cite{Hull:2009zb} we considered the subsector of double field theory in which all fields and parameters are restricted in this way.
Thus we find that
  the strong constraint  used here is equivalent to the  restriction of all fields and parameters to a maximally isotropic  subspace, as stated in~\cite{Hull:2009zb}.

Now, any maximal isotropic subspace of $\mathbb{R}^{2D}$ is related to any other by an $O(D,D)$ transformation.\footnote{
As discussed in e.g.~\cite{Gualtieri}, maximal isotropic subspaces can be associated with pure spinors of $SO(D,D)$. They split into two classes, corresponding to the two chiralities of
$SO(D,D)$ spinors, and any maximal isotropic subspace within a given class is related to any other in that class by an $SO(D,D)$ transformation. Transformations in $O(D,D)$ relate the two classes, so that any maximal isotropic subspace can be obtained from any other by an $O(D,D)$ transformation.
This result can also be seen as a corollary of
 Witt's theorem~\cite{Chevalley,Artin}.}
Then the strong constraint implies that all momenta lie in a totally null subspace that is related by an $O(D,D)$ transformation to the canonical one in which $\tilde p =0$.
As a result,
any subsector of the double field theory satisfying the strong constraint is
related by an $O(D,D)$ transformation to the canonical subsector of fields  with no dependence on the winding coordinates.

Suppose that there are some compactified dimensions, so that
$x^i=(x^\mu, x^a)$ where $x^\mu$ are $n$ coordinates of $\R^{n-1,1}$ and $x^a$ are periodic coordinates of $T^d$. The doubled space then has  further coordinates $\tilde x_i=(\tilde x_\mu, \tilde x_a)$
with $\tilde x_a$ periodic and $\tilde x_\mu$ non-compact.
As discussed in~\cite{Hull:2009zb}, the physically interesting case is that in which the fields are independent of the $\tilde x_\mu$, so that
$\tilde p^\mu=0$.
The toroidal momenta $p_a, \tilde p^a$ are discrete and lie in a discrete  lattice $\Gamma \subset \R^{2d}$, so that the space of allowed momenta is $ \R^n\times \Gamma$.
The subgroup of $O(d,d)$ preserving $\Gamma$ is the T-duality group $O(d,d;\Z)$.
We have seen above that the strong constraint implies that the momenta lie in a totally null subspace
$N$ of $\R^{2D} $ that is related to the canonical one $T^*$ by an $O(D,D)$ transformation.
This $O(D,D)$ transformation is fixed once one chooses a basis for $N$ and a basis for $T^*$ and demands that the $m$'th
basis vector of $N$ is mapped to the $m$'th basis vector of $T^*$.
Here, the momenta actually lie in the space $N'= N\cap ( \R^n\times \Gamma)$,
while the momenta for fields independent of $\tilde x$ are in $T'= T^*\cap ( \R^n\times \Gamma)$.
Choosing the basis vectors for $N$ and $T^*$ to be in $N'$ and $T'$ respectively fixes an
$O(D,D)$ transformation from $N$ to $T^*$ that in fact takes
$N'$ to $T'$ and so must be in the $O(d,d;\Z)$ subgroup of $O(D,D)$.

In summary, the strong constraint (\ref{strongconstr}) implies we can always use the symmetry  $O(D,D)$ or $O(d,d;\Z)$ to  rotate to fields with no dependence on the winding coordinates, which is what we wanted to show.

\section{Reduction to Einstein gravity and derivative expansion}\setcounter{equation}{0}
In this section we are going to check that the action (\ref{THEAction}) or 
(\ref{THEActionMET}) correctly reduces to the standard one for Einstein gravity coupled to 
a two-form and a dilaton
when the dependence on the $\tilde{x}_{i}$ coordinates is dropped. Next, we discuss a derivative expansion
in $\tilde{\partial}^{i}$ and use it to verify the full gauge invariance.

\subsection{Reduction to the Einstein-Kalb-Ramond-dilaton
action}\label{redtotheeinkalramdilact}

In order to recover the standard
action $S_{*}$ in 
(\ref{original})
 we take the action~(\ref{THEActionMET}) and assume that no field depends on 
 the $\tilde x$ coordinates.  Thus,
effectively, we set $\tilde \partial^{i}=0$, which implies
${\cal D}_i = \bar{\cal D}_i = \partial_i$ and
${\cal D}^i = \bar{\cal D}^i = \partial^i= g^{ij} \partial_j$.  The resulting action $\bar S$ takes the form \bea
\label{THEActionvm1}
 \begin{split}
  \bar S  =  \int dx
  e^{-2d}\Big[&
  -\frac{1}{4} \,{g}^{ik}{g}^{jl} {g}^{pq} \Bigl(
    \partial_{p}{\cal E}_{kl}\,  \partial_{q}{\cal E}_{ij}
 - \partial_{i}{\cal E}_{lp} \,  \partial_j{\cal E}_{kq}
  -   \partial_{i}{\cal E}_{pl}\,\partial_j{\cal E}_{qk} \Bigr)
\\ &
 + 2\, \partial^{i}\hskip-1.0pt d~\partial^{j}{g}_{ij}
 +4\,\partial^{i} d \,\partial_{i}d ~\Big]\;,
 \end{split}
 \eea
where  
we drop a constant volume factor  $\int d\tilde{x}$. 
 We now 
 rewrite in terms of $g$ and $b$:
 \begin{equation}
 {\cal E}_{ij} =  g_{ij} + b_{ij}\,.
 \end{equation}
No terms couple derivatives of $g$ and derivatives of $b$.  We find
\bea
\label{THEActionvm2}
 \begin{split}
 \bar S  =  \int dx
  e^{-2d}\Big[&
  -\frac{1}{4} \,g^{ik}g^{jl} g^{pq} \Bigl(
    \partial_{p}g_{kl}\,  \partial_{q}g_{ij}
 -2 \partial_{i}g_{lp} \,  \partial_jg_{kq}
  + \partial_{p}b_{kl}\,  \partial_{q}b_{ij}
 - 2\partial_{i}b_{lp} \,  \partial_jb_{kq}
   \Bigr)
\\ &
 + 2\, \partial^{i}\hskip-1.0pt d~\partial^{j}g_{ij}
 +4\,\partial^{i} d \,\partial_{i}d ~\Big]\;.
 \end{split}
 \eea
Next it will be convenient 
to rewrite the terms involving $b_{ij}$ such that the two-form gauge invariance becomes manifest.
For this we note that the gauge-invariant three-form field strength satisfies
\begin{equation}
\begin{split}
-\frac{1}{12}H^2 &= -\frac{1}{12}g^{ik}g^{jl} g^{pq} H_{ijp}H_{klq}\\
&= -\frac{1}{4} g^{ik}g^{jl} g^{pq} \,\partial_p b_{kl} \,
(\partial_i b_{jq} + \partial_j b_{qi} + \partial_q b_{ij})\\
&= -\frac{1}{4} g^{ik}g^{jl} g^{pq} \bigl(
\,\partial_p b_{kl} \, \partial_q b_{ij}
-2\partial_i b_{lp} \partial_j b_{kq}
 \bigr)\;,
\end{split}
\end{equation}
after relabeling and permuting indices.
Using this
we get
\bea
\label{THEActionvm3}
\bar S  =  \int dx
  e^{-2d}\Big[ \hskip-3pt-\frac{1}{4} g^{ik}g^{jl} \,
    \partial^{p}g_{ij} \partial_{p}g_{kl}\,
  +\frac{1}{2}  g^{pq}  \partial^{i}g_{pj} \,  \partial^jg_{qi}
 + 2\, \partial^{i}\hskip-1.0pt d~\partial^{j}g_{ij}
 +4\,(\partial d)^2   -{1\over 12} H^2\Big]\,.
 \eea

We now consider the action $S_*$, conventionally written as
\begin{equation}
S_*= \int dx \sqrt{-g} e^{-2\phi} \Bigl[ R  + 4 (\partial \phi)^2  -{1\over 12} H^2\Bigr] \,.
\end{equation}
We will show that $S_*$ equals $\bar S$, after a field redefinition
 and   discarding the integral of a total derivative.  The redefinition is
\begin{equation}
\sqrt{-g} e^{-2\phi} = e^{-2d}\,,
\end{equation}
and gives
\begin{equation}
\partial_i \phi =  \partial_i d + {1\over 2} \Gamma_i\,, ~~~~\Gamma_i =
\Gamma^k_{ik} = {1\over 2} g^{kl} \partial_{i}g_{kl} \,.
\end{equation}
Using this in $S_*$ we get
\begin{equation}
S_*= \int dx e^{-2d} \Bigl[ R + g^{ij} \Gamma_i \Gamma_j + 4\,\Gamma_i\partial^id  + 4 (\partial d)^2   -{1\over 12} H^2\Bigr] \,.
\end{equation}
Let us focus on the Einstein term, and integrate by parts the terms with
derivatives of Christoffel symbols,
\begin{equation}
\begin{split}
\int dx e^{-2d} R &=   \int dx e^{-2d}\,  g^{ij} \bigl( \partial_{k}\Gamma^k_{ij}  - \partial_{j}\Gamma^k_{i k}
+ \Gamma^k_{ij} \Gamma_k - \Gamma^l_{ik}
\Gamma^k_{jl} \bigr)  \\
&=  \int dx e^{-2d}\, \Bigl[ 2 \partial^i d ( - \Gamma_i + g^{jk} \Gamma_{ijk} ) +
\partial_j g^{ij} \Gamma_i - \partial_k
 g^{ij} \, \Gamma_{ij}^k \\
  &\hskip60pt  + g^{ij} \bigl(  \Gamma^k_{ij} \Gamma_k - \Gamma^l_{ik}
\Gamma^k_{il}\bigr) \Bigr]\;, \\
\end{split}
\end{equation}
where $ \Gamma_{ijk}=g_{il} \Gamma^l_{jk}=
\frac{1}{2}\left(\partial_{j}g_{ik}+\partial_{k}g_{ji}
-\partial_{i}g_{jk}\right)$.
Further simplifying the terms with the dilaton derivatives yields
\begin{equation}
\begin{split}
\int dx e^{-2d} R =& \int dx e^{-2d}\, \Bigl[ -4\, \partial^i \hskip-1pt d  \,\Gamma_i
+2\,\partial^i \hskip-1pt d  \,\partial^j g_{ij}  +
\partial_j g^{ij} \Gamma_i - \partial_k
 g^{ij} \, \Gamma_{ij}^k \\
  &\hskip60pt  + g^{ij} \bigl(  \Gamma^k_{ij} \Gamma_k - \Gamma^l_{ik}
\Gamma^k_{il}\bigr) \Bigr] \,.
\end{split}
\end{equation}
Using this in $S_*$, we find that the terms coupling $\partial^i d$ to $\Gamma_i$
cancel and we get
\begin{equation}
\begin{split}
S_* =& \int dx e^{-2d}\, \Bigl[~
2\,\partial^i \hskip-1pt d  \,\partial^j g_{ij}
  + 4 (\partial d)^2   -{1\over 12} H^2
  \\&\hskip50pt
+\partial_j g^{ij} \Gamma_i - \partial_k
 g^{ij} \, \Gamma_{ij}^k
   + g^{ij} \bigl(  \Gamma^k_{ij} \Gamma_k - \Gamma^l_{ik}
\Gamma^k_{il}\bigr)+ g^{ij} \Gamma_i \Gamma_j  \Bigr] \,.
\end{split}
\end{equation}
A straightforward computation shows that
\begin{equation}\label{vmbb}
\partial_j g^{ij} \Gamma_{j} - \partial_k
 g^{ij} \, \Gamma_{ij}^k
   + g^{ij} \bigl(  \Gamma^k_{ij} \Gamma_k - \Gamma^l_{ik}
\Gamma^k_{jl} \bigr)+ g^{ij} \Gamma_i \Gamma_j
= -\frac{1}{4} g^{ik}g^{jl} \,
    \partial^{p}g_{ij} \partial_{p}g_{kl}\,
  +\frac{1}{2}  g^{pq}  \partial^{i}g_{pj} \,  \partial^jg_{qi}\,.
\end{equation}
To check this it is best to see first that all structures in (\ref{vmbb})
of the form $g^{ij}\partial_{k}g_{ij}$ cancel out.  Then the left-over terms
combine correctly.  With this
\begin{equation}
\begin{split}
S_* = \int dx e^{-2d}\, \Bigl[
2\,\partial^i \hskip-1pt d  \,\partial^j g_{ij}
  + 4 (\partial d)^2   -{1\over 12} H^2
-\frac{1}{4} g^{ik}g^{jl} \,
    \partial^{p}g_{ij} \partial_{p}g_{kl}\,
  +\frac{1}{2}  g^{pq}  \partial^{i}g_{pj} \,  \partial^jg_{qi} \Bigr]\;.
\end{split}
\end{equation}
This shows that the action $S_*$ is in fact identical, up to total derivatives, 
to the reduced doubled action $\bar S$ in (\ref{THEActionvm3}). 
This is what we wanted to show.

Since the gauge transformations reduce for $\tilde{\partial}^{i}=0$ to the standard gauge transformations, it follows that the reduced
action is gauge invariant. As a warm-up for the proof of the full gauge-invariance it will be instructive to verify this explicitly.
First, we turn to the derivative expansion in $\tilde{\partial}$, which will be useful for organizing the check of gauge invariance.

\subsection{Tilde derivative expansion, dual structure, and gauge  
invariance}\label{dualsection}

Our strategy  here will be to work in a derivative expansion in $\tilde{\partial}$. We write the action (\ref{THEActionMET}) as
 \bea\label{Sexpand}
   S \ = \ S^{(0)}+S^{(1)}+S^{(2)}\;,
 \eea
with the superscript denoting the number of $\tilde{\partial}$ derivatives in the action. In the following we also refer to the corresponding Lagrangians defined by
 \bea
  S^{(k)} \ = \ \int dx d\tilde{x}\,{\cal L}^{(k)}\;, \qquad k=0,1,2\;.
 \eea
Since 
${\cal L}^{(0)}$ contains no $\tilde \partial$ derivatives, 
$S^{(0)}$ takes the same form as
the  action $\bar S$ in equations (\ref{THEActionvm1}) and (\ref{THEActionvm3}), but with fields that depend both on $x^{i}$ and $\tilde{x}_{i}$ (subject to the constraint) and  integration measure  $dxd\tilde{x}$. We thus have the expression, corresponding to
(\ref{THEActionvm1}),
\bea
\label{S0E}
 \begin{split}
  {\cal L}^{(0)} \ = \ {\cal L}^{(0)}\Big[{\cal E},\partial,d\Big] \ = \
  e^{-2d}\Big[&
  -\frac{1}{4} \,{g}^{ik}{g}^{jl} {g}^{pq} \Bigl(
    \partial_{p}{\cal E}_{kl}\,  \partial_{q}{\cal E}_{ij}
 - \partial_{i}{\cal E}_{lp} \,  \partial_j{\cal E}_{kq}
  -   \partial_{i}{\cal E}_{pl}\,\partial_j{\cal E}_{qk} \Bigr)
\\ &
 + 2\, \partial^{i}\hskip-1.0pt d~\partial^{j}{g}_{ij}
 +4\,\partial^{i} d \,\partial_{i}d \Big]\;.
 \end{split}
 \eea
Performing the split ${\cal E}_{ij}=g_{ij}+b_{ij}$, this Lagrangian can be rewritten as
\bea\label{S0}
 \begin{split}
  {\cal L}^{(0)}
  \ = \ e^{-2d}\Big(&-\frac{1}{4}g^{ik}g^{jl}\partial^{p}g_{kl}\,
  \partial_{p}g_{ij}+\frac{1}{2}g^{kl}\partial^{j}g_{ik}\,\partial^{i}g_{jl}
  +2\partial^{i}d\,\partial^{j}g_{ij}+4\partial^{i}d\,\partial_{i}d
  -\frac{1}{12}H^2\Big)\;,
 \end{split}
 \eea
which corresponds to (\ref{THEActionvm3}).

Let us next consider ${\cal L}^{(2)}$. It can be obtained from (\ref{THEActionMET}) by collecting the terms quadratic in $\tilde{\partial}$ from the derivatives
${\cal D}$ and $\bar{\cal D}$. It is given by
 \bea\label{L2}
 \begin{split}
  {\cal L}^{(2)} = e^{-2d}\Big[-&\frac{1}{4}g^{ik}g^{jl}g^{pq}\left({\cal E}_{pr}{\cal E}_{qs}
  \tilde{\partial}^{r}{\cal E}_{kl}\,\tilde{\partial}^{s}{\cal E}_{ij} - {\cal E}_{ir}{\cal E}_{js}
  \tilde{\partial}^{r}{\cal E}_{lp}\,\tilde{\partial}^{s}{\cal E}_{kq}- {\cal E}_{ri}{\cal E}_{sj}
  \tilde{\partial}^{r}{\cal E}_{pl}\,\tilde{\partial}^{s}{\cal E}_{qk}\right) \\
  &-g^{ik}g^{jl}\left({\cal E}_{ip}{\cal E}_{qj}\tilde{\partial}^{p}d\,\tilde{\partial}^{q}{\cal E}_{kl}
  +{\cal E}_{pi}{\cal E}_{jq}\tilde{\partial}^{p}d\,\tilde{\partial}^{q}{\cal E}_{lk}\right)
  +4g^{ij}{\cal E}_{ik}{\cal E}_{jl}\tilde{\partial}^{k}d\,\tilde{\partial}^{l}d\Big]\;.
 \end{split}
 \eea
A crucial observation is that ${\cal L}^{(2)}$ is the T-dual version of ${\cal L}^{(0)}$ in the following sense. Writing ${\cal E}_{ij} = ({\cal E})_{ij}$ with ${\cal E}$ a matrix, we define  the inverse
 \bea\label{fieldred}
  \tilde{{\cal E}}^{ij} \ \equiv \ \left({\cal E}^{-1}\right)_{ij}
  \qquad
  \Rightarrow\qquad  \tilde{\cal E}^{ik}{\cal E}_{kj} \ = \ \delta^{i}{}_{j}\;.
 \eea
The transformation ${\cal E} \to \tilde {\cal E} = {\cal E}^{-1}$
is a special T-duality transformation (an inversion in all circles) with $a=0$, $b=1$, $c=1$, and $d=0$. This implies $M=-{\cal E}$ and
$\bar{M}={\cal E}^{t}$. We can then use (\ref{fsetmat4678599}), with $\tilde{g}_{ij}$
corresponding to $g^{\prime -1}$, to derive
 \bea
  g^{ij} \ = \ \tilde{\cal E}^{ki}\,\tilde{g}_{kl}\,\tilde{\cal E}^{lj}
  \ = \ \tilde{\cal E}^{ik}\,\tilde{g}_{kl}\,\tilde{{\cal E}}^{jl}\;.
 \eea
The inverse of these relations are given by
 \bea\label{gtil}
  \tilde{g}_{kl} \ = \ {\cal E}_{ki}\, g^{ij}\, {\cal E}_{lj}
  \ = \ {\cal E}_{ik}\, g^{ij}\, {\cal E}_{jl}\;.
 \eea

The Lagrangian ${\cal L}^{(2)}$, as given in (\ref{L2}), can be 
obtained from ${\cal L}^{(0)}$ given in (\ref{S0E}) by 
taking 
\bea\label{tildemapping}
  {\cal E}_{ij}\;\rightarrow\; \tilde{\cal E}^{ij}\;, \qquad
  g^{ij}\,\rightarrow\,\tilde{g}_{ij} \;, \qquad
  \partial_{i}\,\rightarrow\; \tilde{\partial}^{i}\;, \qquad
  d\;\rightarrow\; d\;.
 \eea
This is checked by verifying that
\bea
\label{S0Edual}
 \begin{split}
  {\cal L}^{(2)} \ = \
  {\cal L}^{(0)}\Big[ \tilde{\cal E},\tilde{\partial},d\Big] \ = \
  e^{-2d}\Big[&
  -\frac{1}{4} \,{\tilde g}_{ik}{\tilde g}_{jl} {\tilde g}_{pq} \Bigl(
   \tilde \partial^{p} \tilde{\cal E}^{kl}\, \tilde \partial^{q} \tilde{\cal E}^{ij}
 -\tilde \partial^{i} \tilde{\cal E}^{lp} \, \tilde  \partial^j \tilde{\cal E}^{kq}
  - \tilde  \partial^{i} \tilde{\cal E}^{pl}\, \tilde\partial^j \tilde{\cal E}^{qk} \Bigr)
\\ &
 + 2\, \tilde\partial_{i}\hskip-1.0pt d~ \tilde\partial_{j}{\tilde g}^{ij}
 +4\, \tilde \partial^{i} d \, \tilde\partial_{i}d ~\Big]\;.
 \end{split}
 \eea
 Consider, for example, 
the last term in (\ref{L2}). By virtue of (\ref{gtil}) we 
see that it is equal to the last term in (\ref{S0Edual}): 
 \bea
  4e^{-2d}\,{g}^{ij}\,{\cal E}_{ik}\,{\cal E}_{jl}\, \tilde{\partial}^{k}d\,
  \tilde{\partial}^{l}d \ = \
  4e^{-2d}\,\tilde{g}_{ij}\, \tilde{\partial}^{i}d\, \tilde{\partial}^{j}d\;.
 \eea
It is straightforward to check that
all other terms work out similarly, demonstrating
the equality of this expression for ${\cal L}^{(2)}$ with the one given in
(\ref{L2}); this is a consequence of the T-duality invariance.

Next we give the mixed action $S^{(1)}$, which is obtained from (\ref{THEActionMET}) by collecting all terms from the quadratic expressions in ${\cal D}$, $\bar{\cal D}$ that have one ordinary derivative $\partial$ and one tilde derivative $\tilde{\partial}$. The resulting Lagrangian is given by
 \bea\label{S1org}
  \begin{split}
    {\cal L}^{(1)} = e^{-2d}\Big[\,&\frac{1}{2}{g}^{ik}{g}^{jl}{g}^{pq}
    \left({\cal E}_{pr}\,\tilde{\partial}^{r}{\cal E}_{kl}\,\partial_{q}{\cal E}_{ij}
    -{\cal E}_{lr}\,\tilde{\partial}^{r}{\cal E}_{ip}\,
    \partial_{k}{\cal E}_{jq}+{\cal E}_{rl}\,
    \tilde{\partial}^{r}{\cal E}_{pi}\,\partial_{k}{\cal E}_{qj}\right) \\
    &+{g}^{ip}{g}^{jq}\left({\cal E}_{rq}\,\partial_{p}d\,\tilde{\partial}^{r}
    {\cal E}_{ij}-{\cal E}_{pr}\,\tilde{\partial}^{r}d\,\partial_{q}{\cal E}_{ij}
    +{\cal E}_{rp}\,\tilde{\partial}^{r}d\,\partial_{q}{\cal E}_{ij}
    -{\cal E}_{qr}\,\partial_{p}d\,\tilde{\partial}^{r}{\cal E}_{ji}\right)\\
    &-8{g}^{ij}\,{\cal E}_{ik}\,\tilde{\partial}^{k}d\,\partial_{j}d\, \Big] \;.
  \end{split}
 \eea
As before we decompose ${\cal E}_{ij}=g_{ij}+b_{ij}$, after which the Lagrangian reads
 \bea\label{S1gb}
  \begin{split}
   {\cal L}^{(1)} =  e^{-2d}\Big[\,&\frac{1}{2}g^{ik}g^{jl}g^{pq}\left(b_{pr}\,
   \tilde{\partial}^{r}b_{kl}\,\partial_{q}b_{ij}-2b_{lr}\,\tilde{\partial}^{r}b_{ip}\,
   \partial_{k}b_{jq} +b_{pr}\,\tilde{\partial}^{r}g_{kl}\,\partial_{q}g_{ij}
   -2b_{lr}\,\tilde{\partial}^{r}g_{ip}\,\partial_{k}g_{jq}\right)\\
   &+g^{ik}g^{pq}\left(-\tilde{\partial}^{j}g_{ip}\,\partial_{k}b_{jq}-\tilde{\partial}^{j}
   b_{ip}\,\partial_{k}g_{jq}+2b_{rq}\,\partial_{k}d\,\tilde{\partial}^{r}g_{ip}
   -2b_{kr}\,\tilde{\partial}^{r}d\,\partial_{q}g_{ip}\right)\\
   &+2g^{ij}\left(\partial_{j}d\,\tilde{\partial}^{k}b_{ik}-\tilde{\partial}^{k}d\,
   \partial_{j}b_{ki}-4b_{ik}\,\tilde{\partial}^{k}d\,\partial_{j}d\right)\,\Big]\;.
  \end{split}
 \eea
Because of T-duality ${\cal L}^{(1)}$ is invariant under the transformation 
(\ref{tildemapping}).

In order to check the full gauge invariance, we write the gauge 
transformation (\ref{Liegauge}) as
 \bea
  \delta_{\xi} \ = \ \delta_{\xi}^{(0)}+\delta_{\xi}^{(1)}\;,
 \eea
where, again, the superscript denotes  the number of tilde derivatives:
 \bea
  \delta_{\xi}^{(0)}{\cal E}_{ij} 
  &=&  \partial_{i}\tilde{\xi}_{j}
  -\partial_{j}\tilde{\xi}_{i}+{\cal L}_{\xi}{\cal E}_{ij}\;, \\\label{nonlinguage}
  \delta_{\xi}^{(1)}{\cal E}_{ij} 
  &=& -{\cal E}_{ik}\left(\tilde{\partial}^{k}\xi^{l}
  -\tilde{\partial}^{l}\xi^{k}\right){\cal E}_{lj}
  +{\cal L}_{\tilde{\xi}}{\cal E}_{ij}\;,
\eea
using the Lie derivative (\ref{Liegauge}).
Using (\ref{dualvar}) we see that $\delta^{(0)}$ and $\delta^{(1)}$ are T-dual to each other.  
More precisely, under the transformation (\ref{tildemapping}) together with 
$\xi\leftrightarrow \tilde{\xi}$ we have, for instance, 
 \bea
  \delta^{(0)}{\cal E}_{ij}\;\rightarrow\; \delta^{(1)}\tilde{\cal E}^{ij}\;. 
 \eea

Acting on  
(\ref{Sexpand}) with a gauge transformation, we infer that gauge invariance requires
 \bea\label{Svarexp}
 \begin{split}
  &\delta^{(0)}S^{(0)} \ = \ 0\;, \\
  &\delta^{(1)}S^{(2)} \ = \ 0\;, \\
  &\delta^{(0)}S^{(1)}+\delta^{(1)}S^{(0)} \ = \ 0\;,\\
  &\delta^{(1)}S^{(1)}+\delta^{(0)}S^{(2)} \ = \ 0\;.
 \end{split}
 \eea
The first condition is the standard gauge invariance of Einstein's theory coupled to a two-form and a dilaton. The second relation is the T-dual version of this statement, and follows from the first    
by virtue of (\ref{S0Edual}).  
Similarly, the third and fourth relation are T-dual and thus equivalent. Therefore, the only non-trivial check is, say, the third condition, which we verify explicitly.

In general, an action $\int  e^{-2d}L$ constructed from  a density 
 $e^{-2d}$  and a Lagrangian $L$  that transforms as a scalar under
diffeomorphisms is manifestly gauge invariant. More precisely, given 
 \bea
   \delta_{\xi}\big(e^{-2d}\big) \ = \ \partial_{i}\big(\xi^{i}e^{-2d}\big)\;, \qquad 
   \delta_{\xi}L \ = \ \xi^{i}\partial_{i}L\;, 
 \eea
we find   $ \delta_{\xi}\big(e^{-2d}L\big)  =  \partial_{i}\big(\xi^{i}e^{-2d}L\big)$,
and the action is gauge invariant
 (up to a surface term).
 The analogous statements hold for the dual diffeomorphisms
parameterized by $\tilde{\xi}_{i}$.
Therefore, a large part of the variation is guaranteed to combine into total derivatives,
and we only have to keep track of two types of structures during the variation.
First, we have to focus on the terms in the gauge transformations (\ref{finalgt99}) that are not of the form of a Lie derivative, as the non-linear terms in (\ref{nonlinguage}).
Second, we have to compute the variations of terms
that involve a partial derivative, since these
do not transform purely with a Lie derivative. 
An explicit calculation, which we defer to appendix B, shows that all these variations cancel upon
use of the strong form of the constraint 
$\partial_{i}\tilde{\partial}^{i}=0$.  This proves the gauge invariance.

\section{Towards an O(D,D) geometry}\setcounter{equation}{0}

Recall that in the definition (\ref{oddtensor}) of an $O(D,D)$ tensor, position-dependent
$M(X)$ and $\bar M(X)$ matrices control the transformation law.
Due to this $X$-dependence, neither ordinary nor calligraphic derivatives of $O(D,D)$ tensors are $O(D,D)$ tensors.  We introduce here `$O(D,D)$ covariant derivatives'
that acting on $O(D,D)$
tensors
yield $O(D,D)$ tensors.  These derivatives will allow us to write the gauge transformations of ${\cal E}_{ij}$
in  a form similar to that of conventional gravity.
Note that we make no claims for the covariance of our derivatives under gauge transformations.   

We then turn to the construction of an $O(D,D)$ 
scalar ${\cal R}$ that is 
also a scalar
under gauge transformations: 
$\delta_\xi {\cal R} = \xi^M \partial_M {\cal R}$.
This scalar ${\cal R}$
is built from ${\cal E}_{ij}$ and the dilaton $d$ and each term contains two derivatives.
Together with the density $\exp(-2d)$ this scalar can be used to construct an action.
Our investigation shows that this action is equivalent to the earlier one in
(\ref{THEActionMET});  the two differ
by a total derivative term.

Finally, we investigate the relation between T-duality and gauge symmetries by asking to what extent the former can be seen as a special case of the latter.   
We stress that the gauge
transformations are {\it not} diffeomorphisms on the doubled space.
However, we will borrow the language of geometry, and refer to the transformation
$\delta_\xi {\cal R} = \xi^M \partial_M {\cal R}$ as that of a scalar, and refer to covariance and curvatures in a mild abuse of language.

\subsection{O(D,D) covariant derivatives and gauge transformations}
\label{covderodd}

We have already shown in \S\ref{oddcovarianceproven} that the gauge transformations are
$O(D,D)$ covariant.  Since these transformations involve derivatives they
are a natural
place to investigate how to make $O(D,D)$ covariance manifest
by the use of covariant derivatives.
We now introduce $\eta_i$ and $\bar \eta_i$ gauge parameters
similar to the original $\lambda_i$ and $\bar\lambda_i$ in (\ref{parashift}), but
related to 
$\xi^i$ and $\tilde \xi_i$ using
${\cal E}$ instead of $E$:
\be
\eta_i \equiv - \tilde \xi_i  + {\cal E}_{ij} \xi^j \,,~~~~
\bar \eta_i \equiv  \tilde \xi_i + \xi^j {\cal E}_{ji} \,.
\ee
As before, one can also show that
\be
\xi^i = {1\over 2} ( \eta^i + \bar \eta^i) \,,
\ee
where the indices of  $\eta$ and $\bar\eta$ have been raised using  
$g^{-1}$.
 Using the duality transformations (\ref{gfti}) of $\tilde \xi$ and $\xi$,
 it is a few lines  of
 calculation to show that
$\eta$ and $\bar \eta$ are $O(D,D)$ tensors
(with $M=M(X),\bar M = \bar M(X)$ as usual):
\be
 \eta_i (X) =  M_i^{~p}\, \eta'_p (X')\,,~~~~
\bar\eta_i  (X)= \bar M_i^{~p} \,\bar \eta'_p (X')
 \,. \ee
We now rewrite the gauge transformation (\ref{finalgt})  as
 \bea
 \label{finalgt88}
\begin{split}
  \delta {\cal E}_{ij} \ &= \ {\cal D}_i\tilde{\xi}_{j}-\bar{{\cal D}}_{j}\tilde{\xi}_{i}
  +\xi^{M}\partial_{M}{\cal E}_{ij}
+{\cal D}_{i}(\xi^{k}{\cal E}_{kj})-\xi^{k}{\cal D}_{i}{\cal E}_{kj}
+\bar{\cal D}_{j}(\xi^{k}{\cal E}_{ik}) -\xi^{k}\bar{\cal D}_{j}{\cal E}_{ik}\\[0.5ex]
  &= \ {\cal D}_i ( \tilde{\xi}_{j}
+ \xi^{k}{\cal E}_{kj} )
  + \bar{{\cal D}}_{j} (- \tilde{\xi}_{i} +\xi^{k}{\cal E}_{ik})
  +\xi^{M}\partial_{M}{\cal E}_{ij}   -\xi^{k}{\cal D}_{i}{\cal E}_{kj}
 -\xi^{k}\bar{\cal D}_{j}{\cal E}_{ik}\;.
\end{split}
 \eea
The above can be rewritten in terms of $\eta$ and $\bar \eta$ as follows:
\bea
 \label{finalgt87}
\begin{split}
  \delta {\cal E}_{ij} \ &= \ {\cal D}_i \bar\eta_j   + \bar{{\cal D}}_{j}
  \eta_i  +{1\over 2} ( \eta^k {\cal D}_k + \bar \eta^k \bar{\cal D}_k ){\cal E}_{ij}
    -  {1\over 2} ( \eta^k + \bar \eta^k){\cal D}_{i}{\cal E}_{kj}
 -{1\over 2} ( \eta^k+ \bar \eta^k)\bar{\cal D}_{j}{\cal E}_{ik}\;.
\end{split}
 \eea
Reordering we have
\bea
 \label{finalgt87z}
\begin{split}
  \delta {\cal E}_{ij} \ &= \ \underline{{\cal D}_i \bar\eta_j }
    -  {1\over 2} \bigl( \underline{{\cal D}_{i}{\cal E}_{kj}}
    + \bar{\cal D}_{j}{\cal E}_{ik}
      - \bar{\cal D}_k {\cal E}_{ij} \bigr) \, \bar \eta^k
    \\
    &~~  + \underline{\bar{{\cal D}}_{j}  \eta_i }
     -{1\over 2} \bigl( \underline{ \bar{\cal D}_{j}{\cal E}_{ik}}
    + {\cal D}_{i}{\cal E}_{kj}
     - {\cal D}_k {\cal E}_{ij} \bigr) \eta^k \,.
     \end{split}
 \eea
We have underlined
the terms that do not transform
as $O(D,D)$ tensors. In the top line, ${\cal D}_i \bar\eta_j$  does not transform well
because the derivative acts on the $\bar M$ that appears for the transformation of
$\bar \eta$.   All  ${\cal D} {\cal E}$ or
$\bar{\cal D} {\cal E}$ factors are tensors (as we proved earlier),  but the
contraction with $\bar \eta$ on the first line does not respect the index
type:  the $k$ index is unbarred on ${\cal E}$ but barred in $\bar \eta$.
Similar remarks apply to the second line.
We will show below that the non-covariant terms in the variation of the first line and in that of the second line in fact cancel, so that the transformation is covariant.

It is
natural to introduce Christoffel-like symbols and covariant
derivatives.  We define
\be
\begin{split}
~~ ~ {\Gamma}^{\bar k}_{i\bar j} \ &\equiv {1\over 2} \, g^{kl}\left({\cal D}_{i}{\cal E}_{lj}
  +\bar{\cal D}_{j}{\cal E}_{il}-\bar{\cal D}_{l}{\cal E}_{ij}\right)\,,\\
 {\Gamma}^k_{\bar i j} \ & \equiv {1\over 2}\, g^{kl}
   \bigl(\bar{\cal D}_{i}{\cal E}_{jl}
    + {\cal D}_{j}{\cal E}_{li}
     - {\cal D}_l {\cal E}_{ji} \bigr) ~\,.~~~~
 \end{split}
 \ee
With these we can define $O(D,D)$ covariant  derivatives:
\be
\begin{split}
 \nabla_i  \bar \eta_j  &\equiv  {\cal D}_i  \bar \eta_j
 -\Gamma^{\bar k}_{i\bar j}\, \bar \eta_k \,,  \\[0.5ex]
\bar\nabla_j \eta_i & \equiv  \bar{\cal D}_j \eta_i
-  {\Gamma}^k_{\bar ji}\,\eta_k \,.
 \end{split}
 \ee
These are $O(D,D)$ tensors, thus, for example $ \nabla_i  \bar \eta_j$ transforms
as an $O(D,D)$  tensor with an unbarred index $i$ and a barred index $j$.  With the use of the covariant derivatives the  gauge transformations
(\ref{finalgt87z}) become the remarkable
\be
\label{ecovtr}
\delta {\cal E}_{ij} \ = \ \nabla_i \bar\eta_j   + \bar{\nabla}_{j}  \eta_i\,.
\ee
This expression makes manifest the $O(D,D)$ covariance of the gauge transformations.

Additional $O(D,D)$ covariant derivatives are needed for the remaining
index structures.
The following definitions are
$O(D,D)$ covariant:
\begin{equation}
\begin{split}
\nabla_i \eta_j  &=  {\cal D}_i \eta_j  - \Gamma_{ij}^k \, \eta_k \,, ~~~
\Gamma_{ij}^k = {1\over 2} \,g^{kl}  \,{\cal D}_i {\cal E}_{jl}\,, \\
\bar\nabla_i \bar\eta_j  &=  \bar {\cal D}_i \bar \eta_j
 - \Gamma_{\bar i\,\bar j}^{\bar k} \, \bar\eta_k \,, ~~~
\Gamma_{\bar i\, \bar j}^{\bar k} = {1\over 2} \,g^{kl}  \,\bar{\cal D}_i {\cal E}_{lj}\,.
\end{split}
\end{equation}
The covariant derivatives introduced above are metric compatible:
\begin{equation}
\nabla_i g_{jk} = 0\,.
\end{equation}
This is true both if we consider $g$ to have two barred indices or
two un-barred indices.  For two barred indices, for example,  
 we have
\begin{equation}
\begin{split}
\nabla_i g_{\bar j\bar k} &= {\cal D}_i g_{\bar j\bar k} -
\Gamma_{i \bar j}^{\bar l}\, g_{\bar l\bar k} - \Gamma_{i \bar k}^{\bar l}g_{\bar j\bar l} \\
&= {\cal D}_i g_{\bar j\bar k} -
{1\over 2} g^{\bar l\bar p} \bigl( {\cal D}_i {\cal E}_{p\bar j}
+ \bar{\cal D}_j  {\cal E}_{i\bar p}  - \bar {\cal D}_p {\cal E}_{i\bar j} \bigr) g_{\bar l\bar k}
 -{1\over 2}  g^{\bar l\bar p} \bigl( {\cal D}_i {\cal E}_{p\bar k}
+ \bar{\cal D}_k  {\cal E}_{i\bar p}  - \bar {\cal D}_p {\cal E}_{i\bar k} \bigr)
 g_{\bar j\bar l}\\
&= {\cal D}_i g_{\bar j\bar k} -
{1\over 2}  \bigl( {\cal D}_i {\cal E}_{k\bar j}
+ \bar{\cal D}_j  {\cal E}_{i\bar k}  - \bar {\cal D}_k {\cal E}_{i\bar j} \bigr)
 -
{1\over 2}  \bigl( {\cal D}_i {\cal E}_{j\bar k}
+ \bar{\cal D}_k  {\cal E}_{i\bar j}  - \bar {\cal D}_j {\cal E}_{i\bar k} \bigr)
\\
&= {\cal D}_i g_{\bar j\bar k} -
{1\over 2}   {\cal D}_i \bigl({\cal E}_{k\bar j}
+ {\cal E}_{j\bar k} \bigr)    =0\,.
\end{split}
\end{equation}

\bigskip   
The dilaton gauge  transformations (\ref{finalgtINTRO}) can also be 
rewritten in terms of covariant derivatives and the $\eta, \bar \eta$
gauge parameters as follows:
\be
\label{dcovd}
\begin{split}
\delta\hskip1pt d &= \, -{1\over 4} \Bigl(  \, \nabla_i -{1\over 2} 
\bigl( \bar{\cal D}^p {\cal E}_{ip} + 4 {\cal D}_i d\bigr) \Bigr) \, \eta^i 
 -{1\over 4} \Bigl(  \,\bar \nabla_i -{1\over 2} 
\bigl( {\cal D}^p {\cal E}_{pi} + 4 \bar{\cal D}_i d\bigr) \Bigr) \, \bar\eta^i \,.
\end{split}
\ee
In finding this we used $\nabla_i A^j = {\cal D}_i A^j + \Gamma_{ip}^j
A^p$ and a similar equation for barred derivatives.  The above transformations are not as simple as those for ${\cal E}$ 
in (\ref{ecovtr}) and perhaps
indicate that 
defining new connections 
 could be useful.

\bigskip
Let us conclude this section with a verification of the transformation
properties claimed concerning the first line of (\ref{finalgt87z}). The term with the
calligraphic derivative transforms as follows
\be
\label{firsttermvar9}
\begin{split}
{\cal D}_i \bar \eta_j  ~\to ~  M_i^{~k}   \, {\cal D}'_k  \Bigl( \bar M_j^{~p}
\bar \eta'_p \, \Bigr)
 ~=  M_i^{~k}  \,\bar M_j^{~p} \, {\cal D}'_k
\bar \eta'_p  +
\underline{  M_i^{~k}
 ( {\cal D}'_k \bar M_j^{~p}})\bar \eta'_p \;.
\end{split}
\ee
The underlined term
is a non-covariant transformation.
Let us now look at the transformation of the
second underlined term in the first line of (\ref{finalgt87z}):
\be
\label{lckssvm}
 -{1\over 2}\, \bar\eta^k \, {\cal D}_i {\cal E}_{kj}
~\to ~  -{1\over 2} \,\bar\eta'^{p} (\bar M^{-1})_p^{~k}    M_k^{~r}\, M_i^{~s}
\bar M_{j}^{~l} ({\cal D}'_s {\cal E}'_{rl} ) \,.
\ee
This is not covariant because the $k$ indices on $\bar \eta$ and ${\cal E}$
are of different type, thus we get the matrix product $\bar M^{-1}M$
which is not equal to the identity.
We can, however,    easily find the difference from unity:
\be
\bar M^{-1} M =  \bar M^{-1} ( \bar M + (M - \bar M)) = {\bf 1} +
 \bar M^{-1}  (M - \bar M)=  {\bf 1} -2
 \bar M^{-1} g c^t\,,
\ee
where we used (\ref{mbarm}) to evaluate  $M - \bar M$.
Using (\ref{fsetmat4678599})  we find that $ \bar M^{-1}g
=  g'  \bar M^t$  and therefore
\be
\bar M^{-1} M =  {\bf 1} -2 \,
g' \bar M^t c^t = {\bf 1} +2 \, g' c\, M\,,
\ee
where we noted that $cM = - \bar M^t c^t$ (using $cd^t = - dc^t$).
Back in (\ref{lckssvm}) we get
\be
\label{dffdm}
\begin{split}
  -{1\over 2} \bar\eta^k \, {\cal D}_i {\cal E}_{kj}
~\to &~  -{1\over 2} \bar\eta'^{p} ( {\bf 1} +2 \,
g'
 c\, M )_p^{~r}
  \, M_i^{~k} \bar M_{j}^{~l}
\, ({\cal D}'_k {\cal E}'_{rl} )
  \\
=& ~  -{1\over 2} \bar\eta'^{p}  \, M_i^{~k}\bar M_{j}^{~l}
\, ({\cal D}'_k {\cal E}'_{pl} )
-\bar\eta'_{p} ( c\, M )^{pr}\bar M_{j}^{~l}
  \, M_i^{~k}
\, ({\cal D}'_k {\cal E}'_{rl} ) \;.
\end{split}
\ee
The first term on the right-hand side is
covariant, the second is not.
 For that one we use (\ref{pssiu}) in reverse to transform the
${\cal D}'{\cal E}'$ into a ${\cal D}'{\cal E}$:
\be
\label{dffdm99}
\begin{split}
  -{1\over 2} \bar\eta^k \, {\cal D}_i {\cal E}_{kj}
~&\to ~~~  -\,{1\over 2} \bar\eta'^{p}  \, M_i^{~k}\bar M_{j}^{~l}
\, ({\cal D}'_k {\cal E}'_{pl} )
-\bar\eta'_{p} (c\, M )^{pr}
  M_i^{~k} (M^{-1})_r^{~q}
\, ({\cal D}'_k {\cal E}_{qm} ) ({\bar M}^{-1})_l^{~m}  \bar M_{j}^{~l} \\[1.0ex]
&\quad = ~  -{1\over 2} \bar\eta'^{p}  \, M_i^{~k}\bar M_{j}^{~l}
\,( {\cal D}'_k {\cal E}'_{pl} ) -  M_i^{~k}  \bar\eta'_{p}
 \, {\cal D}'_k (c{\cal E} )^p_{~j}
 \\[1.0ex]
&\quad =
~-{1\over 2} \bar\eta'^{p}  \, M_i^{~k}\bar M_{j}^{~l}
\,( {\cal D}'_k {\cal E}'_{pl} )
- M_i^{~k} \bar\eta'_{p}
\, {\cal D}'_k\bar M_j^{~p}\;,
\end{split}
\ee
where we used $\bar M^t =  d + c{\cal E}$.
Finally, making use of (\ref{firsttermvar9})  we see that the two
non-covariant terms
cancel exactly!   As a result, the combination that transforms
covariantly is
\be
{\cal D}_i \bar \eta_j  -{1\over 2} \bar\eta^k \, {\cal D}_i {\cal E}_{kj}
~\to ~M_i^{~k} \bar M_{j}^{~l}  \Bigl[ \,( {\cal D}'_k
\bar \eta'_l )\,   -{1\over 2} \bar\eta'^{p}
\, ({\cal D}'_k {\cal E}'_{pl} )\Bigr]  \,.
\ee
This is what we wanted to show. Note that this is the minimum combination that transforms covariantly.
The  
covariant derivative $\nabla_i$ includes two additional terms that
transform covariantly.

\subsection{Curvature scalar} \label{curvy}

In this section we will construct a curvature ${\cal R}$ that
transforms as a scalar under gauge transformations  
\bea\label{scalartrans}
\delta_\xi {\cal R} =  \xi^M \partial_M  {\cal R} \,,
\eea
 and we will  
show that the background-independent action (\ref{THEAction})
can be written in the Einstein-Hilbert like form
 \bea\label{masteraction}
  S' \ = \ \int dxd\tilde{x}\,e^{-2d}{\cal R}({\cal E},d)\;.   
 \eea
The
 action (\ref{masteraction}) is then clearly  gauge invariant as
$e^{-2d}$ is a density and ${\cal R}$ is a scalar. 
We have seen that
the dilaton is an $O(D,D)$ scalar  
(see  (\ref{dscalar}))
and we will show that ${\cal R}$ is also an $O(D,D)$ scalar.
It will then follow that
 the Lagrangian is an  $O(D,D)$ scalar   
  (as in (\ref{lscalar})) 
 and the action (\ref{masteraction}) is $O(D,D)$ invariant.
 We will refer to ${\cal R}$ as a curvature, even though it does not arise  here from the commutator of covariant derivatives. Indeed, ${\cal R}$ involves the dilaton $d$ while the covariant derivatives $\nabla $
 do not. 
 A similar scalar curvature with a geometric origin was discussed by Siegel \cite{Siegel:1993th}, and it would be interesting to understand better the relation with his curvature.

To get some feeling for the structure of this curvature scalar, we start constructing it perturbatively, as a function ${\cal R} (e,d)$ built
using the fluctuation fields $e_{ij}$ and $d$.
Expanding in fields, we write
\bea
{\cal R}(e,d) =  {\cal R}^{(1)} (e,d)  + {\cal R}^{(2)} (e,d) + \ldots\,,
\eea
where a superscript $(n)$ denotes a term of $n$'th order in fields.
Splitting the gauge transformations similarly, $\delta = \delta^{(0)} +
\delta^{(1)} + \ldots$,
the conditions that ${\cal R}$ is a scalar give
\begin{equation}
\begin{split}
\delta^{(0)} {\cal R}^{(1)}  &~=~ 0\,, \\
 \delta^{(1)}{\cal R}^{(1)}+\delta^{(0)}{\cal R}^{(2)}  &~ = \ \xi^{M}\partial_{M}{\cal R}^{(1)}
  \ = \ \frac{1}{2}\left(\lambda^{i}D_i +\bar{\lambda}^{i}\bar{D}_i\right){\cal R}^{(1)}\;,
\end{split}
\end{equation}
where we used (\ref{parashift}) to express the gauge parameters
in terms of $\lambda, \bar \lambda$.
We
require that the scalar ${\cal R}$ is invariant under the $\mathbb{Z}_2$
transformation
\begin{equation}
e_{ij} \to e_{ji}\;, \, ~~D_i \to \bar D_i \,, ~~\bar D_i \to D_i \,, ~~ d \to d \,.
\end{equation}
This is a symmetry of the action so it
should  be a symmetry of ${\cal R}$.
This symmetry  simplifies the constraint of gauge covariance,
making it sufficient to
check the  transformations with unbarred gauge parameter $\lambda$.
From eqn.~(3.27) of~\cite{Hull:2009mi} we read
\begin{equation}
\label{gaugetrans}
\begin{split}
\delta^{(0)} e_{ij}  &= ~ \bar D_j \lambda_i  \,,  ~~~  ~~
 ~ \delta^{(1)} e_{ij}=  {1\over 2} \Bigl[\,
   (D_i \lambda^k)  e_{kj}
 - \,  (D^k \lambda_i) e_{kj}
   + \,\lambda_k D^k e_{ij} ~\Bigr] \,,   \\
\delta^{(0)} d ~ &=
- {1\over 4}  D\cdot \lambda \,,
~~~~ \delta^{(1)}  d  =
~  {1\over 2}  (\lambda \cdot D) \,d\,~.
 \end{split}
\end{equation}
${\cal R}^{(1)}$ is linear in the fields, has two derivatives, and must be left
invariant by $\delta^{(0)}$.  The unique possibility, up to normalization,
was determined in~\cite{Hull:2009zb}:
\begin{equation}
\label{vmgf}
 {\cal R}^{(1)}(e,d) = \,4 D^2 d +  D^i \bar D^j e_{ij}\,.
\end{equation}
The calculation of ${\cal R}^{(2)}$ uses the strong form of the
constraint $\partial^M \partial_M=0$.
The result is:  \begin{equation}
 \begin{split}
 {\cal R}^{(2)}(e,d) ~=&~ -4 D_i d \,D^i d   -4e^{ij}D_{i}\bar{D}_j d
  -2\, ({D}^{i}e_{ij}\bar D^j d+ \bar{D}^{j}e_{ij}D^i d  )
  \\[0.3ex] &
   -\frac{1}{2}e^{ij} ( D_{i}D^{k}e_{kj} + \bar D_j \bar D^ke_{ik} )
    -\frac{1}{4} ( {D}_le^{li} {D}^k e_{ki} + \bar{D}_le^{il}\bar{D}^k e_{ik} )
    -\frac{1}{4}D^p e^{ij}D_p e_{ij}\;.
 \end{split}
 \end{equation}
Both ${\cal R}^{(1)}$ and ${\cal R}^{(2)}$ are $O(D,D)$
scalars because all index
contractions are of the right kind.
Since the aim of the construction
is a scalar ${\cal R}$ such that (\ref{masteraction}) is the action, we have
verified that
 \bea\label{masteraction99}
  S^{\prime} \ = \ \int dxd\tilde{x}\,e^{-2d}{\cal R} \ = \
  \int dxd\tilde{x}\,\bigl( {\cal R}^{(2)} -2d {\cal R}^{(1)}+\cdots \bigr)\;,
 \eea
 gives an action that to quadratic order in the fluctuations
 reproduces (\ref{redef-action}).

We now extend this to all orders, using the strategy of   \S\ref{Constructingtheaction}.
We find a  background-independent expression that is an $O(D,D)$ scalar that agrees with
 ${\cal R}^{(1)} + {\cal R}^{(2)}$ to quadratic order. We  then argue that this expression is the unique one with these properties and so this must be the desired curvature scalar.
For any term with two derivatives on ${\cal E}$ or on $d$,  $O(D,D)$ covariance requires that the
   second derivative cannot be a ${\cal D}$ but
 must be an $O(D,D)$ covariant derivative $\nabla$, 
 introduced in~\S\ref{covderodd}.
 We maintain the
$\mathbb{Z}_2$ symmetry by introducing symmetrized combinations  where necessary.
The natural background-independent and $O(D,D)$ covariant term that agrees 
with the first term on the right-hand side of (\ref{vmgf}) 
to lowest order then has the following
  expansion to second order in the fields  
   \bea
  2g^{ij}\left(\nabla_{i}{\cal D}_{j}d+\bar{\nabla}_{i}\bar{{\cal D}}_{j}d\right)  = 
  2\left(D^2d+\bar{D}^2d\right)
  -\left(D^{i}e_{ij}\,\bar{D}^{j}d +\bar{D}^{j}e_{ij}\,D^{i}d\right)
  -4e^{ij}D_{i}\bar{D}_{j}d  +  {\cal O} (\hbox{cubic}).
 \eea
Similarly, for the second  term on the right-hand side of (\ref{vmgf})
 we find
 \bea
  \frac{1}{2}g^{ik}g^{jl}\left(\nabla_{k}\bar{\cal D}_{l}{\cal E}_{ij} +
  \bar{\nabla}_{l}{\cal D}_{k}{\cal E}_{ij}\right) &=&
  D^{i}\bar{D}^{j}e_{ij}
  -\frac{1}{4}\Bigl(D^{k}e^{lj}\,D_{l}e_{kj}
  +\bar{D}^{k}e^{jl}\, \bar{D}_{l}e_{jk}\Bigr)
  \\ \nonumber
  &&-\frac{1}{2}e^{ij}\left(D_{i}D^{k}e_{kj}+\bar{D}_{j}\bar{D}^{k}e_{ik}
  \right)+  {\cal O} (\hbox{cubic})\; .
 \eea
In addition to reproducing the terms linear in the fields,  we also obtain 
the   
structures quadratic in the fields that contain an undifferentiated $e_{ij}$. This was necessary for the construction to succeed: such terms  cannot be added in  background-independent form without giving unwanted terms linear in the fluctuations upon expansion about a constant background.  

So far, we have found that
 \bea\label{generalscalar}
  {\cal R}^{(1)} + {\cal R}^{(2)} &=&
  2g^{ij}\left(\nabla_{i}{\cal D}_{j}d+\bar{\nabla}_{i}\bar{{\cal D}}_{j}d\right)
  +\frac{1}{2}g^{ik}g^{jl}\left(\nabla_{k}\bar{\cal D}_{l}{\cal E}_{ij} +
  \bar{\nabla}_{l}{\cal D}_{k}{\cal E}_{ij}\right)\\[0.5ex] \nonumber
  &&+\frac{1}{4}\Bigl(D^{k}e^{lj}\,D_{l}e_{kj}
  +\bar{D}^{k}e^{jl}\, \bar{D}_{l}e_{jk}\Bigr)   -\frac{1}{4} ( {D}_le^{li} {D}^k e_{ki} + \bar{D}_le^{il}\bar{D}^k e_{ik} )\\[0.5ex] \nonumber
  &&-\frac{1}{4}D^p e^{ij}D_p e_{ij}
  -\left(D^{i}e_{ij}\,\bar{D}^{j}d +\bar{D}^{j}e_{ij}\,D^{i}d\right)
  -4D^{i}d\bar{D}_{i}d  +  {\cal O} (\hbox{cubic})\,.
 \eea
There are unique 
background independent expressions that give
the terms on the second and third lines, at least to cubic order.
Replacing the second and third lines with these gives 
 the following candidate for
the full curvature scalar:
  \bea\label{generalscalar}
  {\cal R}({\cal E},d) &=&
  2\left(\nabla^{i}{\cal D}_{i}d+\bar{\nabla}^{i}\bar{{\cal D}}_{i}d\right)
  +\frac{1}{2}\left(\nabla^{i}\bar{\cal D}^{j}{\cal E}_{ij} +
  \bar{\nabla}^{j}{\cal D}^{i}{\cal E}_{ij}\right)\\[0.5ex] \nonumber
  &&+\frac{1}{4}  g^{ij}
 \left(
  {\cal D}^{k}{\cal E}_{lj}\,{\cal D}^{l}{\cal E}_{ki}
  + \bar{{\cal D}}^{k}{\cal E}_{jl}\,\bar{{\cal D}}^{l}{\cal E}_{ik}
  \right)
  -\frac{1}{4}  g^{ij}\left(
  {\cal D}^{l}{\cal E}_{lj}\,{\cal D}^{k}{\cal E}_{ki}
  + \bar{{\cal D}}^{l}{\cal E}_{jl}\,\bar{{\cal D}}^{k}{\cal E}_{ik}
  \right) \\[1.0ex] \nonumber
  &&
  -\frac{1}{4}
  g^{ik}g^{jl}{\cal D}^{p}{\cal E}_{ij}\,{\cal D}_{p}{\cal E}_{kl}
  - \left({\cal D}^{i}{\cal E}_{ij}\, \bar{\cal D}^{j}d
+  \bar{\cal D}^{j}{\cal E}_{ij}\,{\cal D}^{i}d
  \right)
  -4{\cal D}^{i}d\,{\cal D}_{i}d\;.
 \eea
This is 
 background-independent and 
is an $O(D,D)$ scalar
that is second-order in derivatives and which reduces to
${\cal R}^{(1)}+{\cal R}^{(2)}$ upon linearization.
No additional terms can be added that have all these properties:
if they are not to contribute
to quadratic order, 
 they cannot have more than
two ${\cal E}$'s since at least one would be undifferentiated, nor
can they have more than two dilatons since an undifferentiated dilaton
violates the dilaton theorem.  
Then this must be the curvature scalar ${\cal R}$. We now proceed to check that it has the required properties.

The Lagrangian $ {\cal L}$ for our action (\ref{THEActionMET}) differs from $e^{-2d}{\cal R}$ by a total derivative, so that the action $S'$  in  (\ref{masteraction}) is equivalent to  (\ref{THEActionMET}).
Indeed, we show in   appendix C that 
\be
e^{-2d}{\cal R}\, = \,  {\cal L}+\partial_{M}\Theta^{M} \, ,
 \ee
 where  $\Theta^{M}=(\tilde{\theta}_{i},\theta^{i})$ transforms
in the fundamental of $O(D,D)$ so that 
 the last term on the right-hand side is an $O(D,D)$ scalar.
The expressions for $\tilde{\theta}_{i}$ and $\theta^{i}$ are given
in appendix C.  

A rather surprising fact is that the equation of motion of the dilaton --- obtained
by a straightforward 
variation of the action $S$ in
(\ref{THEActionMET}) ---
is precisely  ${\cal R}=0$:
\begin{equation}
\label{dileqn}  \delta S = \int dxd\tilde x\,  e^{-2d}(-2\delta d) \, {\cal R} \,.
\end{equation}
As  $S$ and $S'$ are the same up to total derivatives,
both must give the same equations of motion.   Varying the dilaton in $S'$ gives
\be
\delta S' =  \int dx d\tilde x \, e^{-2d}\,\Bigl[ (-2\delta d) {\cal R}
+ \delta {\cal R} \Bigr]\,.   
\ee
This must agree with the variation in (\ref{dileqn}) and this requires
that  $e^{-2d} \delta {\cal R}$ be a total derivative:  the variation of the dilaton in ${\cal R}$
does not contribute to the equation of motion.  The full equation of motion of the dilaton arises from the variation of $\exp(-2d)$ alone.

Consider the case in which there is no dependence on $\tilde x$, so that  all the terms in ${\cal R}$ involving $ \tilde{\partial}$ drop out. 
  A calculation
similar to that in~\S\ref{redtotheeinkalramdilact} using  $e^{-2d} = \sqrt{-g} e^{-2\phi}$ shows that ${\cal R}$  
reduces to
 \bea\label{Einsteinscalar}
 {\cal R}\,\Big|_{\tilde{\partial}=0} \ = \
  R+4\square\phi-4(\partial\phi)^2-\frac{1}{12}H^2\;.
 \eea
 The Lagrangian $  e^{-2d}\, {\cal R}\,\Big|_{\tilde{\partial}=0}$ defines the action
  \bea\label{masteractionvm}
  S'_*  \ = \ \int dx  
  \sqrt{-g} e^{-2\phi}\Bigl[   R+4\square\phi-4(\partial\phi)^2-\frac{1}{12}H^2
  \Bigr] \,,   
 \eea
which differs by integrals of  total derivatives
from the familiar action $S_*$ in (\ref{original}).  Moreover, one quickly verifies that the dilaton equation of motion arises from $S'_*$ by solely varying the exponential $\exp(-2\phi)$.  This is, of course, the same
equation of motion that follows from $S_*$.

The dilaton equation of motion of $S_*$ can be expressed as a linear
combination of graviton and dilaton beta-functions $\beta_{ij}^g$  
and $\beta^\phi$  of the associated two-dimensional
sigma model. Using the formulae in~\S3.7 of~\cite{Polchinski:1998rq}, 
and $\alpha'=1$ we readily find
that
\be  
 \beta^d \equiv \beta^\phi -{1\over 4} g^{ij} \beta_{ij}^g  
= -{1\over 4} \bigl( 
  R+4\square\phi-4(\partial\phi)^2-\frac{1}{12}H^2 \bigr) \,.
\ee  
Since $e^{-2d} = \sqrt{-g} e^{-2\phi}$, it follows that $\beta^d$, as the
notation suggests,   
can be thought as the beta function for the T-duality invariant dilaton
$d$. Thus, in $S'_*$ the Lagrangian is proportional to the beta function
of the dilaton $d$. 
Indeed,
$\beta^d$ was used
 in~\cite{Tseytlin:1988rr} to 
 write the spacetime action
as $\int \sqrt{-g} e^{-2\phi} \beta^d$.  Let us note that, in contrast,
the integrand
of $S_*$ is not proportional to a linear combination of beta functions.

\subsection{T-duality and gauge invariance}

In this section we examine the relationship between
the $O(D,D)$ duality symmetry
and the gauge symmetry of the theory we have constructed.
We first  consider the theory in $\R^{2D}$ and focus on the infinitesimal form of the
 $O(D,D)$ transformations. We write the group element as ${\bf 1} + T$,
with $T$  in the Lie algebra, and use the basis
 \bea\label{Liebasis}
  \left(\begin{array}{cc} h &
  0 \\ 0 & -h^{t} \end{array}\right)\;, \quad \left(\begin{array}{cc} 0 &
  e \\ 0 & 0 \end{array}\right)\;, \quad \left(\begin{array}{cc} 0 &
  0 \\ f & 0 \end{array}\right)\;,
 \eea
where $h$ is an arbitrary $D\times D$ matrix, while $e$ and $f$ are antisymmetric $D\times D$
matrices.\footnote{We only use $e,f,h$ in this way in this subsection, and this should  not be confused with the uses of $e,f,h$ elsewhere in the paper.}
 From
(\ref{eeprime}), the corresponding infinitesimal $O(D,D)$
transformations are given by
 \bea
 \begin{split}
  h:&\qquad {\cal E}^{\prime}(X^{\prime}) \ = \ {\cal E}(X)+{\cal E}(X)\,h^{t}
  +h\,{\cal E}(X)\;, \\
  e:&\qquad {\cal E}^{\prime}(X^{\prime}) \ = \ {\cal E}(X)+e\;, \\
  f:&\qquad {\cal E}^{\prime}(X^{\prime})  \ = \ {\cal E}(X)-{\cal E}(X)\,f\,{\cal E}(X)\;.
 \end{split}
 \eea
If we write
 \bea\label{xprime}
  X^{\prime M} \ = \ X^{M}-\xi^{M}(X)\;,
 \eea
these transformations can be written as  variations $\delta {\cal E}(X)
\equiv  {\cal E}'(X) - {\cal E}(X)$ that take the form
 \bea\label{oddvar}
  \delta_{h}{\cal E} &=& \xi^{M}\partial_{M}{\cal E}+{\cal E}\,h^t+h\,{\cal E}\;, \\\label{oddvar1}
  \delta_{e}{\cal E} &=& \xi^{M}\partial_{M}{\cal E}+e\;, \\\label{oddvar2}
  \delta_{f}{\cal E} &=& \xi^{M}\partial_{M}{\cal E}-{\cal E}\, f\,{\cal E}\;.
 \eea

The strong constraint implies that the fields are restricted to a null subspace $N$. In the case in which the fields depend just on $x$ and are independent of $\tilde x$, the constraint is satisfied by parameters that depend just on $x$ and are independent of $\tilde x$.
All other choices of $N$ are related to this by an $O(D,D)$ transformation, so it is sufficient to consider this case.
Choosing parameters
\bea\label{Tparameter}
  \tilde{\xi}_{i} \ = \ -\frac{1}{2}e_{ij}x^{j}\;, \qquad
  \xi^{i} \ = \ x^{j}h_j^{~i}\;  ,
 \eea
the gauge transformation (\ref{finalgt})  take
the form
 \bea
  \delta_{\xi}{\cal E}_{ij} \ = \
  \xi^{k}\partial_{k}{\cal E}_{ij}
  +\left({\cal E}\,h^{t}+h\,{\cal E}\right)_{ij}+e_{ij}\;.
 \eea
We see that the $O(D,D)$ transformations
 (\ref{oddvar}) and (\ref{oddvar1}) arise from gauge transformations.
 This is the expected result that the infinitesimal $GL(D,\R)$ transformations with parameter $h$ arise from
 diffeomorphisms and the constant shifts of the $b$-field with parameter $e$ from anti-symmetric tensor gauge transformations.
  Note that exponentiating the $h$ transformations only generates a subgroup of $GL(D,\R)$ (with positive determinant). However, the full  $GL(D,\R)$ symmetry in fact arises from diffeomorphisms
 of the spacetime with coordinates $x^i$.
Thus, the  $GL(D,\R)\ltimes \R^{D(D-1)/2}$ \lq geometric subgroup' of $O(D,D)$  consisting  of
  $GL(D,\R)$ transformations  plus $b$-shifts arises
 from gauge transformations.

 Consider now the remaining infinitesimal $O(D,D)$ transformations with constant parameter $f$.
 A natural ansatz is to consider gauge transformations with parameters
 \bea\label{fparameter}
  \tilde{\xi}_{i} \ = \ 0\;, \qquad
  \xi^{i} \ = \  -{\frac 1 2}f^{ij}\tilde{x}_{j}\;.
 \eea
 The constraint requires that
 $\partial^M \xi^i \partial_M A=0$
for all fields $A$, which here implies
\be
\label{dfgsafads}
f^{ij}\partial _j A \ = \ 0
\ee
for all fields $A$.
In general, the fields ${\cal E}, d$ will depend on all the coordinates $x^i$ and
(\ref{dfgsafads})
will have no solutions, so that the $f$-symmetries do not arise from gauge transformations.
 If the theory, however, is truncated to a subsector in which the fields are also   independent of some of the coordinates $x^i$,   then 
some of 
the $O(D,D)$ transformations generated by $f^{ij}$ 
  do arise from gauge symmetries.
Suppose then that the $D$ coordinates $x^i$ are split into $d$ coordinates $x^a$ and $D-d$ coordinates $x^\mu$, with a corresponding split $\tilde x_i=(\tilde x_\mu, \tilde x_a)$, and consider the truncation to the subsector in which the fields are independent of the $d$ coordinates $x^a$ as well as independent of $\tilde x_i$.
The fields ${\cal E}_{ij}$ and $d$  in the truncated sector then
depend on $x^\mu $ only, so that $\partial _a A=0$ for all fields $A$.
We take  the only non-vanishing components of $f^{ij}$ 
to be
$f^{ab}$, giving  gauge parameters
 \bea\label{fparameterf}
  \tilde{\xi}_{i} \ = \ 0\;, \qquad
  \xi^{\mu} \ = 0
  \;, \qquad
  \xi^{a} \ = \  -{\frac 1 2}f^{ab}\tilde{x}_{b}\;.
 \eea
 The constraint (\ref{dfgsafads}) is now satisfied as $\partial _a A=0$, and so (\ref{fparameterf}) leads to allowed gauge transformations in the truncated subsector. They read
  \bea
  \delta_{\xi}{\cal E}_{ij} \ = \
 -\left({\cal E}\, f\, {\cal E}\right)_{ij}
 = \
 -{\cal E}_{ia} f^{ab} {\cal E}_{bj}\;,
 \eea
 which agrees with (\ref{oddvar2})  using the fact that $\xi^{M}\partial_{M}{\cal E}_{ij}=0$ for these parameters.

 The parameters  $h_a{}^b$ are generators of a $GL(d,\R)$ symmetry 
 arising from diffeomorphisms 
 acting on the $x^a$,
 while the transformations generated by $e_{ab} $ arise from $b$-field gauge transformations.
 Then the parameters $h_a{}^b, e_{ab}, f^{ab}$ 
  are generators of an $O(d,d)$ subgroup of the $O(D,D)$ symmetry.
These infinitesimal $O(d,d)$ symmetries all arise from infinitesimal gauge transformations of the truncated theory, as we have seen. Thus the theory truncated to be independent of the $x^a$ 
(as well as $\tilde x_i$)
has an $O(d,d)$ symmetry that arises from gauge transformations. Strictly speaking, we have only shown this for the subgroup
 that arises from exponentiating infinitesimal generators.
The remaining symmetries in $O(d,d)$ might be thought of as large gauge transformations -- to understand this better would require knowledge  of the finite form of the gauge symmetries.
Note that the vector fields ${\partial \over \partial x^a}$
 can be regarded as $d$ commuting Killing vectors.

 We return now to the general case (without isometries), where the $h$ and $e$ transformations generate a
 $GL(D,\R)\ltimes \R^{D(D-1)/2}$ subgroup of the $O(D,D)$ symmetry that can arise from gauge transformations.
 If $d$ of the $D$ coordinates are compactified on a torus
 the spacetime is $\R^{n-1,1}\times T^d$
and  the $O(d,d)$ subgroup of the $O(D,D)$ symmetry acting on the toroidal directions is broken to $O(d,d;\Z)$ by the periodic boundary conditions.
Then a $GL(d,\Z)\ltimes \Z^{d(d-1)/2}$ subgroup
 of $O(d,d;\Z)$  
 arises through (large) gauge transformations. 
 Indeed, $GL(d,\Z)$
 is the familiar mapping class group of large diffeomorphisms of $T^d$.
In Kaluza-Klein compactification, one
considers  the truncation of the theory to the subsector of fields that are independent of $d$ toroidal coordinates.
As there is no dependence on the $d$ toroidal coordinates, this is the same as the truncation of the theory on $\R^n\times \R^d$ to fields on $\R^n$ considered above.
This truncated theory has a continuous $O(d,d)$ symmetry and we have already seen that  this is generated by gauge transformations on  $\R^n\times \R^d$. However, these are not gauge transformations of  the theory on $\R^n\times T^d$; at most the discrete subgroup $O(d,d;\Z)$ can arise from gauge symmetries of the theory on $\R^n\times T^d$.

The previous discussion sheds an interesting light on the symmetries of Kaluza-Klein theory.
 Any gravity theory reduced on a
$d$-torus has a  $GL(d,\R)$ symmetry  \cite{Cremmer:1979up}. The truncated theory is the same as the theory in which the internal torus is replaced by $\R^d$ with no dependence on the extra $d$ coordinates, and the $GL(d,\R)$ symmetry is then a remnant of the ordinary diffeomorphism symmetry
of $\R^d$. Only a discrete subgroup
$GL(d,\Z)$ is inherited from diffeomorphisms of the internal space $T^d$.
Similarly, the continuous  shift symmetry of the internal two-form is a
remnant of the abelian gauge symmetry of the Kalb-Ramond field on $\R^d$, and again only a discrete subgroup is properly a consequence of the gauge symmetry on $T^d$.
 The most interesting symmetries are the non-linear transformations (denoted by $f$ in (\ref{Liebasis})), which complete the $GL(d)$ and shift symmetries to the duality group $O(d,d)$.
 These do not have a higher-dimensional explanation in the usual formulation and are often referred to as `hidden symmetries', see, e.g.,
\cite{Maharana:1992my,deWit:2000wu} and references therein.
These are now seen as arising from gauge transformations of the double field theory. 
For the Kaluza-Klein theory on
 $\R^n\times T^d$, the
$O(d,d;\Z)$ transformations are precisely the so-called Buscher rules
\cite{Buscher:1987sk,Giveon:1991jj}.
We should stress, however, that at best
only the discrete subgroup $O(d,d;\Z)$ arises from gauge transformation of the theory on
$\R^n\times T^d$, and to
complete the proof of
 this would require knowledge of the so far unknown finite form of the gauge transformations.

\section{Concluding remarks}

In this paper, we have extended spacetime by introducing extra coordinates $\tilde x$ and constructed a theory in the doubled space that 
 is $O(D,D)$  invariant and which reduces to
the conventional theory of Einstein gravity plus antisymmetric tensor plus dilaton when the fields are restricted to be independent of the extra coordinates $\tilde x$.
As such, it could be viewed as an $O(D,D)$ covariantization of the usual theory in which a dependence on dual coordinates  $\tilde x$ is introduced and an $O(D,D)$
invariant constraint   
that effectively removes this dependence is imposed.

 The  theory  has a remarkable  $O(D,D)$ duality symmetry in flat space, broken to a subgroup containing the T-duality group $O(d,d;\Z)$  when $d$ dimensions are  compactified to form a $d$-torus.
Not only are $g$ and $b$ mixed by this symmetry, but the diffeomorphism and
antisymmetric tensor gauge parameters are also rotated into each other.
The T-duality transformations are a generalisation of the usual Buscher transformations to the case in which the fields can have arbitrary dependence on the toroidal coordinates.

The full double field theory that arises from 
the massless sector of closed  
string theory is expected to have many novel features, as we have discussed
elsewhere~\cite{Hull:2009mi,Hull:2009zb}.
Our construction here can be
 viewed as a reduction of
 the complete  double field theory using
our strong constraint, or equivalently restricting to a totally  null subspace.
We view our construction as an important step 
towards the elucidation of the full theory.

An important  
 feature of (\ref{THEActionINTRO})   
  is the background independence. Our construction has been based on a flat doubled space with possible toroidal identifications, but it is interesting to ask
whether it could have a wider validity to more general double spaces.
The action uses double coordinates $X^M$, together with a way of splitting the $2D$  vector indices into two sets of indices, so that a vector
decomposes as $V^M=(\tilde v_i,v^i)$,   
allowing  quantities with $i,j$ indices such as ${\cal E}_{ij}$ to 
be defined.
This  index splitting
 requires that the double space admit an almost local product structure. 
The constraint uses an $O(D,D)$ metric $\eta$.
Here it is a constant metric
and the existence of 
this flat metric  restricts the double space to be a flat space, given by $\R^{2D}$ or obtained from it by suitable identifications.
It would be interesting to seek  generalisations to  more general spaces in which the constraint involves a non-flat metric of signature $(D,D)$.
For example, there is some evidence that doubled groups should play a role in formulating less trivial string backgrounds in this context \cite{Hull:2007jy,Hull:2009sg}.
It would be of considerable interest to seek an extension of this formalism that was applicable to spaces with a torus fibration or to T-folds; in both cases, it is expected that double geometry should play a useful role \cite{Hull:2004in,Hull:2006va}.

It is interesting to discuss the relation of our work to that
in non-symmetric gravity theories~\cite{Damour:1992bt}.
A geometric action for a theory  based on a non-symmetric tensor ${\cal E}$
is usually taken to involve  the use of ${\cal E}$ and its
inverse to define curvatures and torsions and contract indices.
A theory of the type $R+ H^2+...$ that appears in string
theory  requires an infinite number
of geometrical terms and is thus not considered natural in that
framework~\cite{Damour:1992bt}.
Such actions generally have a problem: if  ${\cal E}$  is used to contract indices, the antisymmetric tensor $b$
appears without derivatives, which  violates
gauge invariance.
Our theory avoids this problem by using $g={\frac 12}( {\cal E}+{\cal E}^t)$ to contract indices.

The fields $g$ and $b$ transform independently under diffeomorphisms, so that
 ${\cal E} = g+b$
 transforms reducibly.
 In contrast, the double field theory constructed in this paper admits
an enlarged gauge
symmetry, under which the symmetric and antisymmetric part
of ${\cal E}_{ij}$ transform into each other.
Moreover, the $O(D,D)$ symmetry acts irreducibly on ${\cal E}$. Therefore, the action
(\ref{THEActionINTRO}) can be seen as a geometrical unification of metric and 2-form,
even though it would
not be considered geometrical in the sense of~\cite{Damour:1992bt}
because we contract indices using  the inverse metric $g^{ij}$ instead of the inverse of ${\cal E}_{ij}$.
For us, using $g^{ij}$ is  natural --- it is an $O(D,D)$ tensor and
does not involve $b_{ij}$.
If there had been inverses of ${\cal E}_{ij}$ appearing without derivatives,
there would be   terms with undifferentiated $b_{ij}$ fields which could not be gauge invariant.
A gauge invariant action cannot be constructed using ${\cal E}_{ij}$ alone, and introducing the dilaton $d$ is essential for gauge invariance.

There remain
 a number of issues to be investigated. We have made some preliminary remarks about a possible $O(D,D)$ geometry, but it 
 is important to understand fully  the geometry underlying our theory,
 its symmetries, and 
 the  `scalar curvature' that appears in the  
 action. One geometry has been proposed in this context by
  Siegel \cite{Siegel:1993th}, 
  and generalised geometry~\cite{Tcourant,Hitchin,Gualtieri,LWX,Gualtieri2} might provide a useful geometric framework. 
  It would be interesting to understand whether our construction fits within either of these geometric structures.

There are a number of natural directions in which this research might be developed.
 It would be interesting
to extend the field content to incorporate the RR
$p$-form gauge fields of type II supergravity. 
Another interesting generalisation would be to try to go beyond the T-duality group $O(D,D)$ and
`geometrize'  the exceptional U-duality groups that arise in compactifications of the full type II theory.
In \cite{Hull:2004in}, it was proposed that the double torus representing momentum and string winding modes
 had a natural generalisation to a higher torus representing brane wrapping modes as well, with a natural action of the U-duality group on it %oh
\cite{Hull:1994ys}.  
For toroidal compactifications to 4 dimensions, for example, 
 the appropriate torus is  56-dimensional with a natural action of the U-duality group 
 $E_7(\Z)$~\cite{deWit:2000wu,Hull:2004in}. 
 Generalisations 
of generalised geometry were proposed  in \cite{Hull:2007zu} and developed in \cite{Pacheco:2008ps} in which $O(D,D)$ was replaced by  U-duality groups; 
see also \cite{Hillmann:2009ci}. 
Such geometries may play a key 
 role in U-duality generalisations of our present work.

\subsection*{Acknowledgments}
We are happy to acknowledge helpful correspondence with Marco Gualtieri and discussions with Matthew Headrick,  
Hermann Nicolai, Ashoke Sen, and Dan Waldram.  
This work is supported by the U.S. Department of Energy (DoE) under the cooperative
research agreement DE-FG02-05ER41360. The work of OH is supported by the DFG -- The German Science Foundation.

\appendix

\section{Derivation of the background independent gauge transformations}
\setcounter{equation}{0}

In this appendix, we give some of the details of the derivation of the background independent gauge transformations given in section \ref{BindGaug},
 using results from \cite{Hull:2009zb}.

Recall that the full background independent field is
 \bea
 \label{eroierapp}
{\cal E}_{ij} \equiv E_{ij} + \e_{ij}\;,
 \eea
 with a constant background $E_{ij}$ and a fluctuation field
 $\e_{ij}$   related to the double field theory fluctuation $e_{ij}$
by the matrix relation
\be
\label{mich11app}
\e =  F(e) e =  e F(e)  =  f(e) \,,   
\ee
where the function $F(e)$ is given by 
\be
\label{mich12app}
F(e)  \equiv  \Bigl( 1- {1\over 2} \,e\Bigr)^{-1}   \,.
\ee
It is an immediate consequence of the definitions  that $\e$ and $e$
commute:
\be
\label{mich2}
\e \, e =  e \, \e \,.
\ee
It follows
that
\be
\label{mich3}
e  = \Bigl( 1- {1\over 2} \, e \Bigr) \,\e  = \e\, \Bigl( 1- {1\over 2} \, e \Bigr) \,.
\ee
The above leads to
\be
\label{mich4}
\e - e = {1\over 2} \, e\e  =  {1\over 2}  \e e \,,
\ee
and one readily verifies that
\be
\label{mich5}
\Bigl( 1+ {1\over 2} \, \e \Bigr) \Bigl( 1- {1\over 2} \, e \Bigr)  = 1\,.
\ee
Varying~(\ref{mich4}) and using   (\ref{mich5}) we find
a relation between arbitrary variations,
\be
\label{mich6app}
\delta e   =   \Bigl( 1- {1\over 2} \, e \Bigr) \, \delta \e \,   \Bigl( 1- {1\over 2} \, e \Bigr)\,.
\ee
Then for any variation or derivative
\be
\label{sdfgsdsfgsdg}
\delta {\cal E}= \delta \e= F\delta e F\;.
\ee

It will be useful to split the variation in (\ref{finalgtx})
  into a transport variation
$\delta _T$
given by
 \be
\label{thteegr}
  \delta_T
e_{ij} \ = \    {1\over 2}\,(\lambda \cdot D + \bar \lambda \cdot \bar D) e_{ij}\;,
   \ee
and the rest $ \delta _R$,
so that $\delta e= \delta _T e+ \delta _R e$.
Then
we find
 \be
\label{thteegr}
 \delta_T
 {\cal E} =
F  \delta_T
e F = ~    {1\over 2}\, F\left[ (\lambda \cdot D + \bar \lambda \cdot \bar D) e \right]F
= ~    {1\over 2}\,(\lambda \cdot D + \bar \lambda \cdot \bar D)  {\cal E}\;,
   \ee
   using that (\ref{sdfgsdsfgsdg}) implies
   $D {\cal E}=  F(D e) F$, etc. Thus,
    \be
\label{thteejhkgr}
 \delta_T
 {\cal E}_{ij}
= ~    {1\over 2}\,(\lambda \cdot D + \bar \lambda \cdot \bar D)  {\cal E} _{ij}\;.
   \ee

   Next, we consider
   \be
\label{finalgtiuty}
\begin{split}  \delta_R
e_{ij}  &= ~   \hat D_i \bar\lambda_j +{\hat { \bar D} }_j \lambda_i +{1\over 2} \,   (\hat D_i \lambda^k )\, e_{kj}
+ {1\over 2}
e_{ik} \,
    {\hat { \bar D} }_j \bar \lambda^k\;,
\end{split}
   \ee
which implies
   \be
\label{finalgtiuty00}
\begin{split}  \delta_R
{\cal E}_{kl}  &= ~F_{ki} \left[  \hat D_i \bar\lambda_j +{\hat { \bar D} }_j \lambda_i +{1\over 2} \,   (\hat D_i \lambda^m )\, e_{mj}
+ {1\over 2}
e_{im} \,
    {\hat { \bar D} }_j \bar \lambda^m  \right] F_{jl}\;.
\end{split}
   \ee
We now claim that the derivatives  
 \begin{equation}
 \begin{split}
  {\cal D}_{i}  \ \equiv \  F_{ij}  \hat D_j\;, \qquad
    \bar{{\cal D}}_i  \ \equiv \  F_{ji} {\hat { \bar D} }_j\;,
      \end{split}
 \end{equation}
agree with the previous definitions, as they are given by  
 \begin{equation}
 \begin{split}
  {\cal D}_{i}  \ = \ \partial_i-{\cal E}_{ik}\tilde{\partial}^k\;, \qquad
    \bar{{\cal D}}_i  \ = \ \partial_i+{\cal E}_{ki}\tilde{\partial}^k\;\,.
      \end{split}
 \end{equation}
 This is readily confirmed.  For  example,  
 \bea
  {\cal D} &=&  F(D-\frac  1 2 e \bar D)
  \ = \
  FD-\frac  1 2f \bar D
    \ = \
  (1+\frac  1 2 f) D-\frac  1 2 f\bar D \\ \nonumber
    &=&
  D-\frac  1 2 f(\bar D-D)
   \ = \
  D-   \e \tilde{\partial}
     \ = \
      \partial-{\cal E} \tilde{\partial}\;,
 \eea
where we used $1+\frac  1 2 f=F$.
Then (\ref{finalgtiuty00})
can be rewritten as
  \be
\label{finalgtiuty}
\begin{split}  \delta_R
{\cal E}_{kl}  &= ~    F_{jl} {\cal D}_k \bar\lambda_j + F_{ki} \bar {\cal D}_l \lambda_i +{1\over 2} \,   (  {\cal D}_k \lambda^m )\, \e_{ml}
+ {1\over 2}
\e_{km} \,
\bar  {\cal D}_l \bar \lambda^m\;.
\end{split}
   \ee
Using $ 1+\frac  1 2 f=F$
this can be further rewritten as
  \be
\label{finalgtiuty}
\begin{split}  \delta_R
{\cal E}_{kl}  &= ~    {\cal D}_k \bar\lambda_l +   \bar {\cal D}_l \lambda_k
%\\&
+ {1\over 2} \e_{ml} {\cal D}_k \bar\lambda^m +{1\over 2} \e_{km} \bar {\cal D}_l \lambda^m +{1\over 2} \,   (  {\cal D}_k \lambda^m )\, \e_{ml}
+ {1\over 2}
\e_{km} \,
\bar  {\cal D}_l \bar \lambda^m\;.
\end{split}
   \ee
We then infer
   \be
\label{finalgtiuty}
\delta_R
{\cal E}_{kl}  \ = \    {\cal D}_k \bar\lambda_l +   \bar {\cal D}_l \lambda_k
+ {1\over 2} \e_{ml} {\cal D}_k(  \bar\lambda^m+\lambda^m)  +{1\over 2} \e_{km} \bar {\cal D}_l (\lambda^m+\bar \lambda^m )\;.
\ee

Next we rewrite this in terms of the gauge parameters $(\tilde{\xi},\xi)$.
First, using $ {1\over 2}(\lambda^m+\bar \lambda^m )  = \xi^m$
 we have
  \begin{equation}
  \begin{split}
\label{finalgtiuty}
 \delta_R {\cal E}_{kl}
 \ &= \    {\cal D}_k \bar\lambda_l +   \bar {\cal D}_l \lambda_k
+  ({\cal E}_{ml}    -E_{ml})  {\cal D}_k\xi^m   + ({\cal E}_{km} -E_{km}) \bar {\cal D}_l \xi^m  \\[0.5ex]
 \ &= \    {\cal D}_k( \bar\lambda_l  -E_{ml} \xi^m )+
\bar {\cal D}_l (\lambda_k
-E_{km} \xi^m)
+  {\cal E}_{ml}      {\cal D}_k\xi^m   + {\cal E}_{km}  \bar {\cal D}_l \xi^m\;.
\end{split}
\end{equation}
Using further the inverse relations (\ref{parashift})  
one finds
\be
\begin{split}  
\lambda- E\xi  &= -\tilde{\xi}+E\xi -E\xi =-\tilde{\xi}\;,\\[0.5ex]
\bar \lambda- E^t \xi &= \tilde{\xi}+E^t\xi -E^t\xi =~\tilde{\xi}\;.
\end{split}
\ee
This gives the final result for the gauge transformations
 \bea\label{gaugetr}
  \delta {\cal E}_{ij}
  =
  {\cal D}_i\tilde{\xi}_{j}-\bar{{\cal D}}_{j}\tilde{\xi}_{i}
  +{\cal D}_{i}\xi^{k}{\cal E}_{kj}+\bar{\cal D}_{j}\xi^{k}{\cal E}_{ik}
  +\frac{1}{2}\left(\lambda\cdot D+\bar{\lambda}\cdot \bar{D}\right){\cal E}_{ij}\;.
 \eea
The only terms that look out of place here are those in the last term,
the transport one. They are easily seen, however, to be equal to
$\xi^{i}\partial_{i}+\tilde{\xi}_{i}\tilde{\partial}^{i}
  = \xi^{M}\partial_{M}$.
Then the gauge transformation
can be rewritten as
 \bea
 \label{finalgtapp}
  \delta {\cal E}_{ij} \ = \ {\cal D}_i\tilde{\xi}_{j}-\bar{{\cal D}}_{j}\tilde{\xi}_{i}
  +\xi^{M}\partial_{M}{\cal E}_{ij}
  +{\cal D}_{i}\xi^{k}{\cal E}_{kj}+\bar{\cal D}_{j}\xi^{k}{\cal E}_{ik}\;,
 \eea
which is the final form used in the main text.

\section{Explicit check of gauge invariance}\setcounter{equation}{0}

As a warm-up  we begin by checking that  the action $S^{(0)}$
in (\ref{S0}) 
is gauge invariant under $\delta^{(0)}$.
This, of course, is guaranteed to work. 
As explained in the main text, the strategy will be to keep track only of those transformations that do not take the form of a Lie derivative and that are therefore not guaranteed to yield terms that 
combine into total derivatives. For the variation
$\delta^{(0)}{\cal L}^{(0)}$  
it is therefore sufficient to focus on the terms
that involve a partial derivative.
For the variations of the partial derivatives we use
 \bea\label{noncovvar}
  \delta(\partial_p g_{kl}) &=& {\cal L}_{\xi}(\partial_p g_{kl}) + \partial_{p}\partial_k \xi^{q}
  \,g_{lq}+\partial_p\partial_l\xi^{q}\,g_{kq}\;, \\ \nonumber
  \delta(\partial_i d) &=& {\cal L}_{\xi}(\partial_i d)-\frac{1}{2}\partial_i\partial_j\xi^{j}\;,
 \eea
where the Lie derivative ${\cal L}_{\xi}$
represents the covariant terms.
We do not have to vary the $H^2$ term because the 3-form field strength transforms covariantly.
The variation reads
 \bea
 \begin{split}
  \delta {\cal L}^{(0)} \ = \ e^{-2d}\Big[&-g^{ik}\partial^{p}g_{kl}\,\partial_{p}\partial_i\xi^{l}
  +\partial^{j}g_{ik}\,\partial^{i}\partial_{j}\xi^{k}\\
  &+\partial_{q}g_{ik}\,g^{kl}\partial^{i}\partial_{l}\xi^{q}
  -\partial^{i}\partial_k\xi^{k}\,\partial^{j}g_{ij}\\
  &+2\partial^{i}d\,\partial_{j}\partial_i\xi^{j}+2\partial_id\,g^{kl}\partial_{k}\partial_l\xi^{i}
  -4\partial^{i}d\,\partial_i\partial_j\xi^{j}\Big]\;.
 \end{split}
 \eea
Commuting partial derivatives and relabeling the indices, we see that the two terms in the first line cancel
and that  
two terms in the last line combine.
Since ultimately we have to partially integrate 
it is convenient to rewrite terms in the last line as derivatives on $e^{-2d}$. In total we get
 \bea\label{step}
  \delta{\cal L}^{(0)} &=& e^{-2d}\left(g^{kl}g^{ij}\partial_{q}g_{ik}\,
  \partial_{j}\partial_l\xi^{q}
  -g^{ij}g^{kl}\partial_{j}\partial_{q}\xi^{q}\,\partial_{k}g_{il}\right)\\ \nonumber
  &&+ g^{ij}\partial_{j}\left(e^{-2d}\right)\partial_{i}\partial_q\xi^{q}
  -g^{kl}\partial_i\left(e^{-2d}\right)\partial_{k}\partial_{l}\xi^{i}\;.
 \eea
Here all metrics are written explicitly. If we use
$g^{kl}g^{ij}\partial_{q}g_{ik}=-\partial_{q}g^{lj}$, etc., the terms in the first line can be written up to total derivatives as
 \bea
  -e^{-2d}\partial_{q}g^{lj}\partial_{j}\partial_l\xi^{q}&=&
  \partial_{q}\left(e^{-2d}\right)g^{jl}\partial_{j}\partial_{l}\xi^{q}
  +e^{-2d}g^{lj}\partial_{l}\partial_j\partial_q\xi^q\;, \\ \nonumber
  e^{-2d}\partial_k g^{jk}\partial_{j}\partial_q\xi^{q} &=&
  -\partial_{k}\left(e^{-2d}\right)g^{jk}\partial_j\partial_q\xi^q
  -e^{-2d}g^{jk}\partial_k\partial_j\partial_q\xi^q\;.
 \eea
The terms with $\partial^3\xi$ cancel each other, while the terms with
$\partial\left(e^{-2d}\right)$ cancel against the terms in the second line of (\ref{step}).
Thus, as expected, $\delta{\cal L}^{(0)}=0$ follows.

We now turn to the proof of full gauge invariance  
using the derivative expansion (\ref{Sexpand}).
As we stated in the main text,  
it is sufficient to verify  
 the third condition in (\ref{Svarexp}).
For this, 
 it is convenient to  simplify the expression for $S^{(1)}$
 given in (\ref{S1gb}).
Upon relabeling of indices, the terms cubic in $b$ can be rewritten in a form that is proportional to the invariant field strength $H_{ijk}$.
To identify terms unambiguously 
we rewrite every term with a single dilaton derivative term according to
$e^{-2d}\partial_{i}d=-\tfrac{1}{2}\partial_{i}(e^{-2d})$ and then partially integrate in order to move the derivative away from the dilaton. This leads to further simplifications and the final form of ${\cal L}^{(1)}$ reads
 \bea\label{S1final}
  \begin{split}
    {\cal L}^{(1)} =  e^{-2d}\Big[\,&\frac{1}{2}g^{ik}g^{jl}g^{pq}\left(b_{ir}\,
    \tilde{\partial}^{r}b_{jp}\,H_{klq}+b_{pr}\,\tilde{\partial}^{r}g_{kl}\,\partial_{q}g_{ij}
    -2b_{lr}\,\tilde{\partial}^{r}g_{ip}\,\partial_{k}g_{jq}\right) \\
    &-g^{ik}g^{pq}\,\tilde{\partial}^{j}b_{ip}\,\partial_{k}g_{jq}
    +2b_{ir}\,\partial_{j}\tilde{\partial}^{r}g^{ij}
    +2\tilde{\partial}^{k}b_{ik}\,\partial_{j}g^{ij}\\
    &+2g^{ij}\partial_{i}\tilde{\partial}^{k}b_{jk}
    +\tilde{\partial}^{k}g^{ij}\,\partial_{i}b_{jk}-8g^{ij}\,b_{ik}\,
    \tilde{\partial}^{k}d\,\partial_{j}d\;\Big]+({\rm td})\;.
  \end{split}
 \eea
We do the same with the single dilaton derivative
in ${\cal L}^{(0)}$,
 \bea
  \begin{split}
   {\cal L}^{(0)} \ = \ e^{-2d}\Big(&-\frac{1}{4}g^{ik}g^{jl}\partial^{p}g_{kl}\,
   \partial_{p}g_{ij}+\frac{1}{2}g^{kl}\partial^{j}g_{ik}\,\partial^{i}g_{jl}
   \\&-\partial_{i}\partial_{j}g^{ij}+4\partial^{i}d\,\partial_{i}d
   -\frac{1}{12}H^{ijk}H_{ijk}\Big)+({\rm td})\;.
  \end{split}
 \eea

For simplicity we begin the check of gauge invariance with only
$\tilde{\xi}_{i}$ non-zero.  
To vary ${\cal L}^{(1)}$ we use  
$\delta^{(0)}b_{ij}=\partial_{i}\tilde{\xi}_{j}-\partial_{j}\tilde{\xi}_{i}$, while
$\delta^{(0)}$ is trivial on all other fields.  
Thus we have to vary only
the $b$'s inside (\ref{S1final}). For the variation of
${\cal L}^{(0)}$ under $\delta^{(1)}$ we use  
 \begin{eqnarray}
  \delta^{(1)}g_{ij} \ =  {\cal L}_{\tilde{\xi}}g_{ij}\;, \qquad
  \delta^{(1)}b_{ij} \ = \ {\cal L}_{\tilde{\xi}}b_{ij}\;, \qquad
  \delta^{(1)}d \ = \ \tilde{\xi}_{i}\tilde{\partial}^{i}d-\frac{1}{2}\tilde{\partial}^{i}
  \tilde{\xi}_{i}\;,
 \end{eqnarray}
where ${\cal L}_{\tilde{\xi}}$ denote the Lie derivative with respect to the `dual diffeomorphism' parameter $\tilde{\xi}_{i}$. Moreover, the dilaton transforms as a density under these dual diffeomorphisms.
Again, 
it is guaranteed that all variations that are covariant in this sense combine
into the total derivative $\tilde{\partial}^{i}(\tilde{\xi}_{i}{\cal L})$.
Since we 
have an integration over $d\tilde{x}$, these total derivatives can be ignored.
We only have to work out the variations that are non-covariant (in the dual sense), and the only source for those terms are partial derivatives. Even the $H$ field 
 is not covariant anymore under the dual diffeomorphisms:
 \bea
  \delta_{\tilde{\xi}}^{(1)}H_{ijk} \ = \ {\cal L}_{\tilde{\xi}}H_{ijk}
  +3\tilde{\partial}^{p}\tilde{\xi}_{[i}\,\partial_{|p|}b_{jk]}+3\partial_{[i}\tilde{\xi}_{|p|}\,\tilde{\partial}^{p}b_{jk]}
  +6\partial_{[i}\tilde{\partial}^{p}\tilde{\xi}_{j}\,b_{k]p}\;.
 \eea
For the partial derivatives of $g_{ij}$, $b_{ij}$, and $d$ one finds
 \begin{eqnarray}
   \delta_{\tilde{\xi}}^{(1)}\big(\partial_{i}g_{jk}\big) &=& {\cal L}_{\tilde{\xi}}
   \big(\partial_{i}g_{jk}\big)+\tilde{\partial}^{p}\tilde{\xi}_{i}\,\partial_{p}g_{jk}
   +\partial_{i}\tilde{\xi}_{p}\,\tilde{\partial}^{p}g_{jk}
   -2\partial_{i}\tilde{\partial}^{p}\tilde{\xi}_{(j}g_{k)p}\;, \\
   \delta_{\tilde{\xi}}^{(1)}\big(\partial_{i}b_{jk}\big) &=&
   {\cal L}_{\tilde{\xi}}\big(\partial_{i}b_{jk}\big)
   +\tilde{\partial}^{p}\tilde{\xi}_{i}\,\partial_{p}b_{jk}
   +\partial_{i}\tilde{\xi}_{p}\,
   \tilde{\partial}^{p}b_{jk}+2\partial_{i}\tilde{\partial}^{p}\tilde{\xi}_{[j}\,b_{k]p}\;,\\
   \delta_{\tilde{\xi}}^{(1)}\big(\partial_{i}g^{jk}\big) &=& {\cal L}_{\tilde{\xi}}
   \big(\partial_{i}g^{jk}\big)
   +\tilde{\partial}^{p}\tilde{\xi}_{i}\,\partial_{p}g^{jk}
   +\partial_{i}\tilde{\xi}_{p}\,\tilde{\partial}^{p}g^{jk}
   +2g^{p(j}\partial_{i}\tilde{\partial}^{k)}\tilde{\xi}_{p}\;, \\
   \delta_{\tilde{\xi}}^{(1)}\big(\partial_{i}d\big) &=& {\cal L}_{\tilde{\xi}}
   \big(\partial_{i}d\big)+\tilde{\partial}^{k}\tilde{\xi}_{i}\,\partial_{k}d+\partial_{i}
   \tilde{\xi}_{k}\,\tilde{\partial}^{k}d-\frac{1}{2}\partial_{i}\tilde{\partial}^{k}
   \tilde{\xi}_{k}\;.
 \end{eqnarray}
Finally, we need the variation of the double (partial) divergence of $g^{ij}$,
 \begin{eqnarray}
  \delta_{\tilde{\xi}}^{(1)}\big(\partial_{i}\partial_{j}g^{ij}\big) =
  {\cal L}_{\tilde{\xi}}\big(\partial_{i}\partial_{j}g^{ij}\big)
  +2\partial_{i}\partial_{j}g^{ip}\,
  \tilde{\partial}^{j}\tilde{\xi}_{p}  
  +2\partial_{i}\tilde{\xi}_{p}\,\tilde{\partial}^{p}
  \partial_{j}g^{ij} +2\partial_{j}g^{ip}\,\partial_{i}\tilde{\partial}^{j}\tilde{\xi}_{p}
  +\partial_{i}\partial_{j}\tilde{\xi}_{p}\,\tilde{\partial}^{p}g^{ij}\;.
 \end{eqnarray}
It is now straightforward to vary the partial derivatives in
${\cal L}^{(0)}$ by the non-covariant terms given here  
and the $b_{ij}$ in ${\cal L}^{(1)}$ 
according to the standard abelian 2-form transformations.
To see how this gauge invariance works let us illustrate the cancellation for the terms quadratic in derivatives on $d$. Varying these terms in
${\cal L}^{(0)}$ and ${\cal L}^{(1)}$ one finds  
 \bea
  \delta^{(0)}{\cal L}^{(1)}
  +\delta^{(1)}{\cal L}^{(0)}
  &=& \delta^{(0)}\left(-8e^{-2d}g^{ij}b_{ik}\tilde{\partial}^{k}d\,\partial_{j}d\right)
  +\delta^{(1)}\left(4e^{-2d}g^{ij}\partial_{i}d\,\partial_{j}d\right)\\ \nonumber
  &=& 8e^{-2d}\partial^{i}d\left(\partial_{k}\tilde{\xi}_{i}\,\tilde{\partial}^{k}d
  -\partial_{i}\tilde{\xi}_{k}\,\tilde{\partial}^{k}d
  +\tilde{\partial}^{k}\tilde{\xi}_{i}\,\partial_{k}d
  +\partial_{i}\tilde{\xi}_{k}\,\tilde{\partial}^{k}d
  -\frac{1}{2}\partial_{i}
  \tilde{\partial}^{j}\tilde{\xi}_{j}\right) \\ \nonumber
  &=& 8e^{-2d}\partial^{i}d\left(\partial_{k}\tilde{\xi}_{i}\,\tilde{\partial}^{k}d
  +\tilde{\partial}^{k}\tilde{\xi}_{i}\,\partial_{k}d\right)
  -4e^{-2d}\partial^{i}d\,\partial_{i}\tilde{\partial}^{k}\tilde{\xi}_{k} \\ \nonumber
  &=& -2e^{-2d}\left(\partial_{j}g^{ij}\,\partial_{i}\tilde{\partial}^{k}\tilde{\xi}_{k}
  +g^{ij}\partial_{i}\partial_{j}\tilde{\partial}^{k}\tilde{\xi}_{k}\right)  \;,
 \eea
where in the last equation we used the constraint and performed a partial integration.
As stated, all terms quadratic in $d$ have cancelled. The remaining structures cancel against other contributions.
Indeed, one may check that all remaining variations cancel 
without performing
partial integrations  since 
we have fixed the possible total-derivative ambiguity by moving all derivatives away from the dilaton. This proves full gauge invariance under
$\tilde{\xi}_{i}$.

Let us now turn to gauge invariance under $\xi^{i}$. As before, we split
${\cal E}_{ij}=g_{ij}+b_{ij}$. With only $\xi^{i}$ non-vanishing $\delta^{(0)}$ takes the form of the standard diffeomorphism symmetry,
 \bea
  \delta_{\xi}^{(0)}g_{ij} \ = \ {\cal L}_{\xi}g_{ij}\;, \qquad
  \delta_{\xi}^{(0)}b_{ij} \ = \ {\cal L}_{\xi}b_{ij}\;, \qquad
  \delta_{\xi}^{(0)}d \ = \ \xi^{i}\partial_{i}d-\frac{1}{2}\partial_{i}\xi^{i}\;,
 \eea
with the dilaton transforming as a density. Thus, in $S^{(1)}$ we only have to vary the terms that involve partial derivatives and therefore transform non-covariantly. For this we use (\ref{noncovvar}) together with
 \bea\label{bignoncov1}
  \delta_{\xi}^{(0)}\big(\tilde{\partial}^{r}b_{jp}\big) &=&
  {\cal L}_{\xi} \big(\tilde{\partial}^{r}b_{jp}\big)
  +\tilde{\partial}^{k}b_{jp}\,\partial_{k}\xi^{r}
  +\tilde{\partial}^{r}\xi^{k}
  \partial_{k}b_{jp}-2\tilde{\partial}^{r}\partial_{[j}\xi^{k}\,b_{p]k}\;, \\
  \delta_{\xi}^{(0)}\big(\tilde{\partial}^{k}g^{ij}\big) &=&
  {\cal L}_{\xi}\big(\tilde{\partial}^{k}g^{ij}\big)
  +\tilde{\partial}^{p}g^{ij}\,\partial_{p}\xi^{k}
  +\tilde{\partial}^{k}\xi^{p}\,
  \partial_{p}g^{ij}-2\tilde{\partial}^{k}\partial_{p}\xi^{(i}\, g^{j)p}\;, \\
  \delta_{\xi}^{(0)}\big(\tilde{\partial}^{k}d\big) &=& {\cal L}_{\xi}
  \big(\tilde{\partial}^{k}d\big)
  +\tilde{\partial}^{p}d\,\partial_{p}\xi^{k}
  +\tilde{\partial}^{k}\xi^{j}\,\partial_{j}d-\frac{1}{2}
  \tilde{\partial}^{k}\partial_{j}\xi^{j}\;, \\
  \delta_{\xi}^{(0)}\big(\tilde{\partial}^{r}g_{kl}\big) &=&
  {\cal L}_{\xi}\big(\tilde{\partial}^{r}g_{kl}\big)
  +\tilde{\partial}^{p}g_{kl}\,\partial_{p}\xi^{r}
  +\tilde{\partial}^{r}\xi^{p}\,
  \partial_{p}g_{kl}+2\tilde{\partial}^{r}\partial_{(k}\xi^{p}\,g_{l)p}\;, \\
  \delta_{\xi}^{(0)}\big(\partial_{i}b_{jk}\big) &=&
  {\cal L}_{\xi}\big(\partial_{i}b_{jk}\big) -2\partial_{i}\partial_{[j}\xi^{p}\,b_{k]p}\;,\\
  \delta_{\xi}^{(0)}\big(\tilde{\partial}^{r}\partial_{j}g^{ij}\big) &=&
  {\cal L}_{\xi}\big(\tilde{\partial}^{r}\partial_{j}g^{ij}\big)
  +\tilde{\partial}^{p}\partial_{j}g^{ij}\,\partial_{p}\xi^{r}
  +\tilde{\partial}^{r}\xi^{p}\,\partial_{p}\partial_{j}g^{ij}-\partial_{j}g^{pj}\,
  \tilde{\partial}^{r}\partial_{p}\xi^{i}\\ \nonumber
  &&-\tilde{\partial}^{r}g^{pj}\,\partial_{j}
  \partial_{p}\xi^{i}
  -g^{pj}\tilde{\partial}^{r}\partial_{j}\partial_{p}\xi^{i}
  -\tilde{\partial}^{r}g^{ip}\,\partial_{p}\partial_{j}\xi^{j}
  -g^{ip}\tilde{\partial}^{r}\partial_{p}\partial_{j}\xi^{j}\;, \\ \label{bignoncov2}
  \delta_{\xi}^{(0)}\big(\partial_{i}\tilde{\partial}^{k}b_{jk}\big) &=&
  {\cal L}_{\xi}\big(\partial_{i}\tilde{\partial}^{k}b_{jk}\big)
  +\partial_{i}\partial_{k}\xi^{p}\,\tilde{\partial}^{k}b_{jp}
  +\partial_{k}\xi^{p}\,\partial_{i}\tilde{\partial}^{k}b_{jp}
  +\partial_{i}\partial_{j}\xi^{p}\,\tilde{\partial}^{k}b_{pk}\\ \nonumber
  &&+\partial_{i}\tilde{\partial}^{k}\xi^{p}\,\partial_{p}b_{jk}
  +\tilde{\partial}^{k}\xi^{p}\,\partial_{i}\partial_{p}b_{jk}
  +\partial_{i}\tilde{\partial}^{k}\partial_{j}\xi^{p}\,b_{pk}
  -\tilde{\partial}^{k}\partial_{j}\xi^{p}\,\partial_{i}b_{kp}\;.
 \eea
Next, we look at $\delta^{(1)}$. It acts trivially on $d$, while on $g$ and $b$ we find the non-linear transformations
 \bea\label{nonlingauge1}
  \delta_{\xi}^{(1)}g_{ij} &=& 2\big(\tilde{\partial}^{k}\xi^{l}-\tilde{\partial}^{l}
  \xi^{k}\big)g_{k(i}\,b_{j)l}\;,\\
  \delta_{\xi}^{(1)}g^{ij} &=& -\big(\tilde{\partial}^{i}\xi^{k}
  -\tilde{\partial}^{k}\xi^{i}\big)g^{jl}b_{lk}+\left(i\leftrightarrow j\right)\;,\\ \label{nonlingauge2}
  \delta_{\xi}^{(1)}b_{ij} &=& g_{ik}\big(\tilde{\partial}^{l}\xi^{k}-\tilde{\partial}^{k}
  \xi^{l}\big)g_{lj}+b_{ik}\big(\tilde{\partial}^{l}\xi^{k}-\tilde{\partial}^{k}
  \xi^{l}\big)b_{lj}\;.
 \eea
The variation $\delta^{(1)}{\cal L}^{(0)}$ is lengthy because
we have to vary the metric everywhere, not only under partial derivatives. For this it is convenient to slightly rewrite ${\cal L}^{(0)}$ with less appearances of metrics and inverse metrics,
 \bea
  \begin{split}  
   {\cal L}^{(0)} =  e^{-2d}\Big(&\frac{1}{4}g^{pq}\partial_{p}g^{ij}\,
   \partial_{q}g_{ij}-\frac{1}{2}g^{ij}\partial_{j}g^{kl}\,\partial_{l}g_{ik}
   -\partial_{i}\partial_{j}g^{ij}+4g^{ij}\partial_{i}d\,\partial_{j}d
   -\frac{1}{12}g^{il}g^{jp}g^{kq}H_{ijk}H_{lpq}\Big)\;.
  \end{split}
 \eea
Here we 
need to vary everything under $\delta^{(1)}$ as given in (\ref{nonlingauge1})\,--\,(\ref{nonlingauge2}).
Then we have to vary all terms involving partial derivatives in (\ref{S1final}) according to
(\ref{noncovvar}) and (\ref{bignoncov1})\,--\,(\ref{bignoncov2}). A tedious but straightforward calculation then shows that these two sets of variations precisely cancel, thus proving gauge invariance.

\section{Properties of the curvature scalar}

In section \S \ref{curvy} we defined a curvature scalar ${\cal R}$ and claimed that it transforms as (\ref{scalartrans}), and that the
 action (\ref{masteraction}) is equivalent to (\ref{THEAction}). We also claimed that the dilaton equation of motion is precisely ${\cal R}=0$. In this appendix we verify these claims.

We first test these ideas for  $S^{(0)}$.  We  add
a total derivative
to the corresponding  
Lagrangian density  $\mathcal{L}^{(0)}$ in (\ref{S0}), 
so that the dilaton equation arises just by varying the exponential prefactor.  Then we confirm that the resulting Lagrangian density
${\mathcal{L}'}^{(0)}$
agrees with ${\cal R}$, also evaluated to leading order in
the tilde-derivative expansion.  Indeed, 
choosing
the total derivative term
\be\label{totder}
{\mathcal{L}'}^{(0)} \equiv ~\mathcal{L}^{(0)}+\partial_i \Bigl[  e^{-2d} \, g^{ij} ( 4\partial_j d  + \partial^{k} g_{jk})\Bigr]  ,
\ee
leads to
\bea\label{S099}
 \begin{split}
{\mathcal{L}'}^{(0)}
=~&e^{-2d}\Big[
  -\frac{1}{4}g^{ik}g^{jl}\partial^{p}g_{kl}\,
  \partial_{p}g_{ij}+\frac{1}{2}g^{kl}\partial^{j}g_{ik}\,\partial^{i}g_{jl}
  + \partial_i \bigl( g^{ij} \partial^k g_{jk} \bigr)  -\frac{1}{12}H^2
  \\[0.5ex] &
  \hskip25pt +4\bigl( - \partial^{i}d\,\partial^{j}g_{ij}-\partial^{i}d\,\partial_{i}d
  + g^{ij} \partial_i \partial_j d\bigr)
\Big] \;. \end{split}
 \eea
It is
 straighforward to confirm that the variation of $d$ in the
terms on the last line of the above equation yields a total derivative.
This confirms that the dilaton equation of motion that results from
${\mathcal{L}'}^{(0)}$ is obtained by varying only the dilaton
exponential; the equation of motion is simply the vanishing of the
terms within 
square
brackets.

 It is natural to expect that ${\mathcal{L}'}^{(0)}$
is the Lagrangian associated with $S'$, when tilde-derivatives are set
to zero.   To confirm this we evaluate ${\cal R}$ for $\tilde \partial=0$.  A  
computation
starting with (\ref{generalscalar}) gives
 \bea\label{generalscalar99}
 {\cal R}\Big|_{\tilde{\partial}=0} &=&
  2\left(g^{ip} \Bigl[ \partial_p - {1\over 2} \partial^k {\cal E}_{kp} \Bigr] \partial_{i}d
  +g^{ip}  \Bigl[ \partial_p - {1\over 2} \partial^k {\cal E}_{pk} \Bigr] \partial_{i}d\right) \nonumber \\[0.5ex]
  &&
  +\frac{1}{2}\left(g^{ip} \Bigl[ \partial_p
  - {1\over 2} \partial^k {\cal E}_{kp} \Bigr] \partial^{j}{\cal E}_{ij} +
  g^{ip} \Bigl[ \partial_p
   - {1\over 2} \partial^k {\cal E}_{pk} \Bigr]
   \partial^{j}{\cal E}_{ji}\right)  
   \\[0.5ex] \nonumber
  &&+\frac{1}{4}  g^{ij}
 \left(
 \partial^{k}{\cal E}_{lj}\,\partial^{l}{\cal E}_{ki}
  + \partial^{k}{\cal E}_{jl}\,\partial^{l}{\cal E}_{ik}
  \right)
  -\frac{1}{4}  g^{ij}\left(
  \partial^{l}{\cal E}_{lj}\,\partial^{k}{\cal E}_{ki}
  + \partial^{l}{\cal E}_{jl}\,\partial^{k}{\cal E}_{ik}
  \right) \\[1.0ex] \nonumber
  &&
  -\frac{1}{4}
  g^{ik}g^{jl}\partial^{p}{\cal E}_{kl}\,\partial_{p}{\cal E}_{ij}
  - 2
 \partial^{j}g_{ij}\,\partial^{i}d
  -4\partial^{i}d\,\partial_{i}d\;.
 \eea
Collecting terms this becomes  
\bea\label{generalscalar99}
 {\cal R}\Big|_{\tilde{\partial}=0} &=&
4\bigl( - \partial^{i}d\,\partial^{j}g_{ij}-\partial^{i}d\,\partial_{i}d
  + g^{ij} \partial_i \partial_j d\bigr)
 \nonumber \\[0.5ex]
  &&+\frac{1}{4}  g^{ij}
 \left(
 \partial^{k}{\cal E}_{lj}\,\partial^{l}{\cal E}_{ki}
  + \partial^{k}{\cal E}_{jl}\,\partial^{l}{\cal E}_{ik}
  \right)
  -\frac{1}{4}  g^{ij}\left(
  \partial^{l}{\cal E}_{lj}\,\partial^{k}{\cal E}_{ki}
  + \partial^{l}{\cal E}_{jl}\,\partial^{k}{\cal E}_{ik}
  \right) \\[1.0ex] \nonumber
  &&
 -\frac{1}{4}  g^{ij}\left(
  \partial^{k}{\cal E}_{kj}\,\partial^{l}{\cal E}_{il}
  + \partial^{k}{\cal E}_{jk}\,\partial^{l}{\cal E}_{li}
  \right)  -\frac{1}{4}
  g^{ik}g^{jl}\partial^{p}{\cal E}_{kl}\,\partial_{p}{\cal E}_{ij}
  + \partial^i \partial^j  g_{ij} \;.
 \eea
A short calculation then shows that, as expected,   
\be
\label{lpreqr}
{\mathcal{L}'}^{(0)} = e^{-2d}  {\cal R}\Big|_{\tilde{\partial}=0} \,.
\ee

So far we have proved that $e^{-2d}{\cal R}$ and the original Lagrangian ${\cal L}$ differ 
only by a total derivative when restricted to $\tilde{\partial}=0$.  Indeed, combining
(\ref{totder}) and (\ref{lpreqr}) we get
  \bea
  e^{-2d}{\cal R}\Big|_{\tilde{\partial}=0} \ = \ {\cal L} \Big|_{\tilde{\partial}=0}
  +  \partial_i \Bigl[  e^{-2d} \, g^{ij} ( 4\partial_j d  + \partial^{k} g_{jk})\Bigr]\,.
 \eea
We now state a simple but useful lemma.  Given two $O(D,D)$ scalars
$A (x, \tilde x)$ and $B(x, \tilde x)$ that differ by   total derivative terms,
then after an $O(D,D)$ transformation 
 they again differ by  total derivative terms.
The proof is immediate.  
Let the scalars $A,B$ 
differ by 
total derivative terms
in the $(x, \tilde x)$ frame, 
\be
A(x, \tilde x) =  B (x, \tilde x)  +  \partial_i  F^i  + \tilde\partial^i  \tilde F_i\,.
\ee
Being $O(D,D)$ scalars we have $A(x, \tilde x) = A'(x' ,\tilde x')$
and $B(x, \tilde x) = B'(x' ,\tilde x')$, so that  the above becomes
\be
\label{modtodr}
A'(x', \tilde x') =  B' (x', \tilde x')  +  \partial_i  F^i  + \tilde\partial^i  \tilde F_i\,.
\ee
Recall now  that under an $O(D,D)$ transformation (\ref{odddefstr}) 
the derivatives  
$\partial^M = ( \partial_i \,, \tilde \partial^i)$ 
transform as
\be
 \begin{pmatrix} \partial' \\[0.5ex] \tilde {\partial'}\end{pmatrix} =  
\begin{pmatrix}     a& b\\[0.5ex] c & d\end{pmatrix}
\begin{pmatrix} \partial \\[0.5ex] \tilde \partial\end{pmatrix}
\quad \to \quad  
\begin{pmatrix} \partial \\[0.5ex] \tilde \partial\end{pmatrix}
=  \begin{pmatrix} d^t& b^t \\[0.5ex] c^t& a^t \end{pmatrix}
\begin{pmatrix} \partial' \\[0.5ex] \tilde {\partial'}\end{pmatrix} \,.
\ee
Since  $a,b,c,d$ are constant matrices, the total derivative terms in (\ref{modtodr}) remain  
total derivatives in the primed variables. This proves the lemma.

Returning to our application, consider the two $O(D,D)$ scalars: ${\cal L}$
and $e^{-2d} {\cal R}$.  Given the strong constraint,
 one can always use an $O(D,D)$ transformation to 
  rotate into a frame where fields have no $\tilde x$ dependence,  allowing
us to set $\tilde{\partial}=0$.  In this frame we have verified that the two scalars differ by a total derivative.  The lemma implies
 that  the 
 original Lagrangians ${\cal L}$ and  $e^{-2d} {\cal R}$ 
  differed by a total derivative before the $O(D,D)$ transformation.

We can describe the general form of the total derivatives
more explicitly by writing
\be
e^{-2d}{\cal R}\, = \,  {\cal L}+\partial_{M}\Theta^{M} \,.
 \ee
The last term on the right-hand side must be an $O(D,D)$ scalar since the other two  terms are.
This happens  
 if  $\Theta^{M}=(\tilde{\theta}_{i},\theta^{i})$ transforms
in the fundamental of $O(D,D)$.
The components of $\Theta^M$ can be constructed~as
 \bea\label{thetadefa}
  \theta^{i} \ = \ \frac{1}{2}\left(-Y^{i}+\bar{Y}^{i}\right)\;, \qquad  
  \tilde{\theta}_{i} \ = \ \frac{1}{2}\left({\cal E}_{ji}Y^{j}
  +{\cal E}_{ij}\bar{Y}^{j}\right)\;,
 \eea
where $Y$ and $\bar{Y}$, to be determined below, 
 transform with $M$ and $\bar{M}$
and are therefore $O(D,D)$ tensors in the sense of 
\S\ref{oddcovarianceproven}. This ensures  
 that  $\Theta^{M}$ transforms in the fundamental of $O(D,D)$. This statement should be compared with equations (\ref{CalInv}) that relate $\partial_{M}=(\tilde{\partial}^{i},\partial_{i})$ transforming in the fundamental to ${\cal D}$ and $\bar{\cal D}$ transforming with $M$ and
$\bar{M}$, respectively. For these variables the proof was given
in \S4.2 of \cite{Hull:2009mi}, and this proof readily extends to any $\Theta^{M}$ defined as in
(\ref{thetadefa}). Specifically, here we have
 \bea
  Y^{i} &=& -e^{-2d}g^{ij}\left(4{\cal D}_{j}d+g^{kl}\bar{\cal D}_{k}{\cal E}_{jl}\right)\;,
  \\
  \bar{Y}^{i} &=&
 ~ \, e^{-2d}g^{ij}\left(4\bar{\cal D}_{j}d+g^{kl}{\cal D}_{k}{\cal E}_{lj}\right)\;.
 \eea
These expressions are fixed by the requirement that they transform covariantly
under $O(D,D)$, i.e., with
$M$ and $\bar{M}$, and that  $\theta^{i}$ correctly reduces to
(\ref{totder}) for $\tilde{\partial}=0$. As a consistency check one may verify that
$\tilde{\theta}_{i}$ reduces for $\partial=0$ to the T-dual expression, the one which is obtained from (\ref{totder}) by mapping
${\cal E}\rightarrow \tilde{\cal E}$, $\partial\rightarrow \tilde{\partial}$
as in \S\ref{dualsection}.

\bigskip
We 
now  relate ${\mathcal{L}'}^{(0)}$ or $ {\cal R}\big|_{\tilde{\partial}=0}$ to the standard quantities in the conventional action.  A calculation
similar to that in~\S\ref{redtotheeinkalramdilact} gives
 \bea\label{Einsteinscalara}
 {\cal R}\,\Big|_{\tilde{\partial}=0} \ = \
  R+4\square\phi-4(\partial\phi)^2-\frac{1}{12}H^2\;.
 \eea
 This means that to zeroth-order in the tilde derivative expansion we have
  \bea\label{masteractionvm}
  S'_*  \ = \ \int dx  
  \sqrt{-g} e^{-2\phi}\Bigl[   R+4\square\phi-4(\partial\phi)^2-\frac{1}{12}H^2
  \Bigr] \,.   
 \eea
Equation (\ref{Einsteinscalara}) also 
gives further evidence 
that ${\cal R}$ is a scalar under gauge transformations.  We use an $O(D,D)$ transformation to fields with no $\tilde x $ dependence so that
 ${\cal R}$ takes the above form, and the gauge transformations become
 the familiar ones for which the right-hand side is clearly a scalar.

\end{document}